\documentclass[10pt]{article}

\usepackage{a4wide}
\usepackage[margin=2cm]{geometry}
\usepackage{amsmath}
\usepackage{amssymb}
\usepackage{amsthm}
\usepackage[mathscr]{euscript}
\usepackage{graphicx}
\usepackage{wrapfig}
\usepackage{caption}
\usepackage{subcaption}
\usepackage{booktabs}
\usepackage[bookmarks=true,bookmarksopen=true,colorlinks=true,breaklinks=true,linkcolor=blue,citecolor=blue]{hyperref}

\newcommand{\Rc}{R_{\mathrm{circ}}}

\newcommand{\dz}{\,\mathrm{d}z}
\newcommand{\dr}{\,\mathrm{d}\rho}

\newcommand {\md} {\mathrm d}

\numberwithin{equation}{section}

\newcommand{\Eqref}[1]{Eq.~\eqref{#1}}

\newcommand{\Sectionref}[1]{Section~\ref{#1}}

\newcommand{\Figref}[1]{Figure~\ref{#1}}

\begin{document}

\title{Cosmic String and Black Hole Limits of Toroidal Vlasov Bodies in General Relativity}
\author{
Ellery Ames,
\thanks{Department of Mathematics, KTH, S-10044 Stockholm Sweden (\texttt{ellery@kth.se})}
\quad
H\aa kan Andr\'easson,
\thanks{
Department of Mathematical Sciences,
Chalmers University of Technology and University of Gothenburg, 
S-41296 G\"oteborg, Sweden} 
\thanks{(\texttt{hand@chalmers.se})}
\quad
Anders Logg
\footnotemark[2]
\thanks{(\texttt{logg@chalmers.se})}
}
\date{\today}
\maketitle

\begin{abstract}
We numerically investigate limits of a two-parameter family of stationary solutions to the Einstein-Vlasov system. The solutions are toroidal and have non-vanishing angular momentum. As one tunes to more relativistic solutions (measured for example by an increasing redshift) there exists a sequence of solutions which approaches the extreme Kerr black hole family. Solutions with angular momentum larger than the square of the mass are also investigated, and in the relativistic limit the near-field geometry of such solutions is observed to become conical in the sense that there is a deficit angle. Such solutions may provide self-consistent models for rotating circular cosmic strings.
\end{abstract}

\section{Introduction}
In the previous work \cite{Ames:2016vu} three different types of stationary solutions to the axially symmetric Einstein-Vlasov system were constructed numerically. These were disc-like solutions, spindle-like solutions and toroidal solutions. The main aim of the study \cite{Ames:2016vu} was to go beyond the analytic solutions which were obtained in \cite{Andreasson:2011hg} and \cite{Andreasson:ch} as perturbations of spherically symmetric Newtonian solutions. In particular, a question that was raised in \cite{Andreasson:ch} was if there exist regular stationary solutions which contain ergoregions. This question was answered affirmatively in \cite{Ames:2016vu} where it was found that the most relativistic members of the family of toroidal solutions do contain ergoregions.

The presence of ergoregions suggests that one may be approaching the family of Kerr black hole solutions, and one aim of the present study is to investigate whether one has a sequence of stationary solutions which have Kerr-family limiting members. Such a \emph{quasistationary transition} to black hole solutions does not occur in the spherically symmetric setting due to a Buchdahl bound $2 \mathcal M/ \mathcal R < 8/9$, which applies to large classes of matter models, cf. \cite{Andreasson:2008fu}. In this case there is thus a gap such that $2 \mathcal M/ \mathcal R$ cannot approach one. However, if one allows for charge a similar bound relating the mass, radius, and total charge is known \cite{Andreasson:2008ge}, and in this case there is no gap; that is a quasistationary transition to an extremal Reissner-Nordstr\"om black hole could be possible. Indeed, that this is the case has been shown in \cite{Meinel:cf}.
In the case where one has angular momentum the model black hole solutions are the Kerr family, which are parametrized by the mass $\mathcal M$ and angular momentum $\mathcal J$ with the restriction $|\mathcal J| \le \mathcal M^2$. Here equality is achieved for the extremal Kerr solution. In this case a quasistationary transition of fluid bodies to extremal black holes has been shown in the case of disc solutions for dust analytically by Meinel \cite{Meinel:2006eh,Meinel:2004hj}, and studied in more general cases numerically by Ansorg et al. \cite{Meinel:2012tn,Ansorg:2003dk,Fischer:2005bw}. In particular, their study includes families of toroidal bodies. Below, we provide evidence of a quasistationary transition to an extreme Kerr black hole for a class of rotating toroidal solutions to the Einstein-Vlasov system.

To this end we study a two-parameter family of solutions parametrized by $E_0$ and $L_0$. The black hole limit is approached by decreasing $E_0$ for a certain critical value of $L_0$. This critical solution sequence has the feature that the extremal black hole solution is approached from the stable side of the binding energy curve and while $|\mathcal J| \ge \mathcal M^2$. We are however, not able to go all the way to the extreme black hole limit. At some point near the black hole limit the code halts the approach to the black hole and instead finds a different, and apparently more stable, solution path towards a distinct limit. Solutions on this path resemble classical (non-quantum) rotating circular cosmic strings.

The solutions on the cosmic string path have angular momentum which is larger than $\mathcal M^2$, and thus, due to angular momentum conservation, are stable against collapse to a Kerr black hole. The properties of these solutions are investigated and it is discovered that such solutions have conical geometry in the near-field regime suggestive of cosmic strings. Conical geometry is characterized by a non-vanishing deficit angle. The deficit angles that we compute for the extremely thin toroidal solutions along the ``cosmic string path" agree with results that have been derived for models where the matter is represented by Dirac type sources and where the spacetime geometry is computed for such a given singular matter distribution. Studies by Garfinkle and coauthors for self-consistent solutions of the Einstein-scalar-gauge field system in the case of straight cosmic strings \cite{Garfinkle:1985ui,Futamase:1988iu,Garfinkle:1989gw} indicate that such agreement is reasonable for small deficit angles and weak fields, but that in general the deficit angle depends on the specifics of the matter model.

Why is it that large angular momentum solution sequences change course from approaching a black hole limit to a string limit? We guess that sufficiently close to the black hole the solutions become unstable and there is evidence that our algorithm, which is based on a fixed-point iteration scheme, ``prefers" stable solutions. In spherical symmetry the numerical study \cite{Andreasson:2006dza} provides a characterization of stable and unstable static solutions of the Einstein-Vlasov system. Using this characterization we have investigated which static solutions can be found by our algorithm. The conclusion is, roughly, that the algorithm only converges in the case of stable solutions. Hence in order to get closer to the black hole limit it seems necessary to use a different numerical scheme.

Before presenting the outline of the paper let us briefly comment on some previous numerical studies of the static and stationary Einstein-Vlasov system. In spherical symmetry the static Einstein-Vlasov system has the advantage that it is sufficient to solve an initial value problem for an ODE to construct solutions. Hence, there is no difficulty concerning convergence as both stable and unstable solutions are easily constructed. Nevertheless, the structure and properties of these solutions are quite interesting, cf. \cite{Andreasson:2007ix} and \cite{Andreasson:2011dza}. In addition, in the spherically symmetric case there exist static \textit{massless} solutions, cf. \cite{Andreasson:2016jo} and \cite{Akbarian:2014gt}. Earlier results are due to Shapiro and Teukolsky (and collaborators) who investigated the Einstein-Vlasov system numerically in a long series of papers spanning from 1987 to 1994. Their main aim was to study the evolution problem in the axially symmetric case but in their general program they also constructed stationary axially symmetric solutions in \cite{Shapiro:1993hi,Shapiro:1993gb}.

The outline of the paper is as follows. In \Sectionref{sec.EVSystem} we review the formulation of the Einstein-Vlasov system, referring to \cite{Ames:2016vu} for more details. The results of solving a boundary value problem for this system are presented in \Sectionref{sec.Results}. In particular we discuss the overall behavior of the parameter space (\Sectionref{sec.Results.Overview}), as well as the black hole (\Sectionref{sec.BHlimit}) and cosmic string (\Sectionref{sec.CosmicStringLimit}) limits. We end the paper with a brief discussion and conclusions in \Sectionref{sec.DiscussionConlusions}.

\section{The Axisymmetric Einstein-Vlasov System}
\label{sec.EVSystem}
The formulation and numerical solution of the equations closely follows that in \cite{Ames:2016vu}, to which we refer the reader for details.
We parametrize the metric as in Bardeen \cite{Bardeen:1973ux}
\begin{equation}
\label{eq:Metric}
g = - e^{2 \nu} dt^2 + e^{2 \mu} d\rho^2 + e^{2 \mu}dz^2 + \rho^2 B^2 e^{-2 \nu} (d\varphi- \omega dt)^2,
\end{equation}
where the coordinates $(t, \varphi)$ are associated to the time and angular commuting Killing fields respectfully, and the metric fields $\nu, \mu, \omega, B$ depend only on $\rho \in [0, \infty)$ and $z \in (-\infty, \infty)$. Vlasov matter is modeled by a distribution function $f$, depending on spacetime coordinates $x = (t,\rho, z, \varphi)$ and four-momenta $p$. The momenta are taken to lie in the mass-shell $\mathcal P$, defined at each spacetime point $x$ as the subset of forward oriented vectors satisfying $g_x(p,p) = -m^2$, where $m$ is the particle rest mass. The particle mass is assumed the same for all particles, and we make the choice $m=1$. The distribution function is transported along the geodesic flow of the spacetime by the Vlasov equation, and the coupled Einstein-Vlasov system is closed through an energy momentum tensor which takes the form
\begin{equation}
\label{eq.VlasovEMTensor}
T_{ij}(x) : = \int_{\mathbb R^3} p_i p_j f(x, p) \sqrt{-\det g} \, \frac{\md p^1\md p^2\md p^3}{-p_0}.
\end{equation}
To solve the coupled Einstein-Vlasov system we make an ansatz that the Vlasov distribution depends on the phase-space coordinates $(x,p)$ only through the particle angular momentum
\begin{align}
\label{eq:AngMomentumExpression}
\begin{split}
L &= (\rho B)^2 e^{-2 \nu} ( p^3 - \omega p^0)  \\
   &= \rho B e^{-\nu} v^3,
\end{split}
\end{align}
and particle energy
\begin{align}
\label{eq:EnergyExpression}
\begin{split}
E &= e^{2 \nu} p^0 + \omega (\rho B)^2 e^{-2\nu}  (p^3 - \omega p^0) \\
& = e^\nu \sqrt{m^2 + \sum_{i = 1}^3 (v^i)^2 } + \omega L,
\end{split}
\end{align}
where it is convenient to work in the frame $v^0 = e^\nu p^0,  v^1 = e^\mu p^1, v^2 = e^\mu p^2, v^3 = \rho B e^{-\nu} ( p^3  - \omega p^0)$.
In general this ansatz takes the form $f(x, p) = F(E,L)$.
Since $E$ and $L$ are conserved along the geodesics traveled by the particles, this choice of distribution function ensures that the Vlasov equation is satisfied. As a result, the full Einstein-Vlasov system is reduced to an elliptic integro-differential system of equations for the metric fields,
\begin{align}
\label{eq:EinsteEqNU}
\Delta \nu & =
4 \pi \left(
\Phi_{00} + \Phi_{11}
+ \left( 1 + (\rho B)^2 e^{-4 \nu} \omega^2 \right)\Phi_{33}
+ 2 e^{-4 \nu} \omega \Phi_{03} \right)  \\\nonumber
& - \frac 1B \nabla B \cdot \nabla \nu
+ \frac 12 e^{-4\nu} (\rho B)^2 \nabla \omega \cdot \nabla \omega,  \\
\Delta B & = 8 \pi B \Phi_{11}  - \frac 1\rho \nabla \rho \cdot \nabla B, \\
\Delta \mu & =
- 4 \pi \left(
\Phi_{00} + \Phi_{11}
+ \left((\rho B)^2 e^{-4 \nu} \omega^2 - 1 \right)\Phi_{33}
+ 2 e^{-4 \nu} \omega \Phi_{03} \right)  \\ \nonumber
& + \frac 1B \nabla B \cdot \nabla \nu - \nabla \nu \cdot \nabla \nu
+ \frac 1\rho \nabla \rho \cdot \nabla \mu + \frac 1\rho \nabla \rho \cdot \nabla \nu
+ \frac 14 e^{-4\nu} (\rho B)^2 \nabla \omega \cdot \nabla \omega,   \\
\label{eq:EinsteEqWW}
\Delta \omega & = \frac{16 \pi}{(\rho B)^2} \left( \Phi_{03} + (\rho B)^2 \omega \Phi_{33} \right)
- \frac 3B  \nabla B \cdot \nabla \omega + 4 \nabla \nu \cdot \nabla \omega
- \frac 2\rho \nabla \rho \cdot \nabla \omega,
\end{align}
where $\Delta u:= \rho^{-1} \partial_\rho(\rho^{-1} \partial_\rho u) + \partial_z\partial_z u$ and $\nabla u = (\partial_\rho u, \partial_z u)$. The variables $\Phi_{ij}$ represent convenient combinations of the energy momentum integrals given by
\begin{align}
\label{eq.PHIs}
\Phi_{00} 	=  e^{2\mu - 2 \nu} T_{tt} ,
\quad
\Phi_{11} 	= T_{\rho \rho} + T_{zz},
\quad
\Phi_{33} 	= (\rho B)^{-2} e^{2\mu + 2 \nu} T_{\varphi \varphi},
\quad
\Phi_{03} 	=   e^{2\mu + 2 \nu} T_{t \varphi}.
\end{align}

For the present work we use a generalized polytropic ansatz given by
\begin{equation}
\label{eq.ansatz}
F(E, L) = A (E_0-E)_+^k (L-L_0)_+^l,
\end{equation}
for an amplitude $A$, and where $(\cdot)_+$ indicates that only the positive part is taken. Furthermore, we make the ``democratic choice'' $k = l = 0$, meaning that all particle energies (resp. particle angular momenta) are equally weighted. In this ansatz the parameter $E_0$ specifies the maximum energy of a particle in the body, while the parameter $L_0$ specifies the minimum particle angular momentum. For any choice of  parameters $E_0, L_0$ and a given metric, the amplitude $A$ is fixed by taking the solution to have total mass $\mathcal M$. In our simulations $\mathcal M$ is taken to be one. The result is a two-parameter family of ansatzes parametrized by $E_0, L_0$.

In any axisymmetric and stationary solution, the total mass $\mathcal M$ and angular momentum $\mathcal J$ can be computed via Komar integrals \cite{PhysRev.113.934}, for which we obtain
\begin{align}
\mathcal M & = 2\pi \int_{z = -\infty}^\infty \int_{\rho = 0}^\infty B\left(\Phi_{00} + \Phi_{11} + \Phi_{33}(1 - (\rho B)^2\omega^2 e^{-4\nu})\right)\rho \dr \dz, \\
\mathcal J & =  -2\pi\int_{z = -\infty}^\infty \int_{\rho = 0}^\infty e^{-4 \nu}B\left(\Phi_{03} + \omega (\rho B)^2\Phi_{33}\right) \rho \dr \dz.
\end{align}
When comparing our regular solutions to black hole solutions, it is useful to distinguish the cases $\mathcal M^2 /|\mathcal J|$ is greater or less than one. For Kerr black hole solutions, $\mathcal M^2 /|\mathcal J| \ge 1$, but for solutions with sufficiently large angular momentum no black holes can form. The special case of $\mathcal M^2 /|\mathcal J| = 1$ is the (family of) extremal Kerr black holes.

A quantity of interest is the binding energy (or mass-defect), which specifies how much energy would be released upon forming the gravitationally bound body (cf. \cite{Zeldovich:gx}). Below in \Sectionref{sec.Results} we give values for the fractional binding energy $E_b = (\mathcal M_0 - \mathcal M) / \mathcal M_0$, which is the ratio of the binding energy to the total rest mass $\mathcal M_0$, where
\begin{equation}
\mathcal M_0 = 2\pi \int_{z = -\infty}^\infty \int_{\rho = 0}^\infty B e^{2\mu} \left( \int_{\mathcal P_x} f(x, p) p^0 \frac{\sqrt{-\det g(x)}}{- p_0} \md p^1 \md p^2 \md p^3\right) \rho \dr \dz.
\end{equation}

The system \Eqref{eq:EinsteEqNU}--\Eqref{eq:EinsteEqWW} is solved numerically by an adaptive finite element method implemented with FEniCS~\cite{Logg:2012jw,Logg:2010kt}, using a code built on the one described in \cite{Ames:2016vu}. The main addition is an adaptive mesh refinement scheme which is necessary in order to resolve the extremely dense configurations while maintaining appropriate asymptotic boundary conditions. Our mesh refinement scheme uses error indicators constructed from the jump in the normal derivative of $\nu$ across the cell boundaries and implements D\"orfler marking \cite{dorfler_convergent_1996} with a $30 \%$ marking fraction. We have also implemented an Anderson
\cite{anderson1965iterative,walker2011anderson} acceleration scheme with variable depth (set to 5 for extreme solutions presented in this article), which improves the convergence and stability of the fixed-point iteration. Details will be presented elsewhere.

\section{Results}
\label{sec.Results}
\subsection{Overview}
\label{sec.Results.Overview}
\begin{figure}[h!]
    \centering
    \begin{subfigure}[t]{0.48\textwidth}
        \centering
	\includegraphics[height=4.5cm]{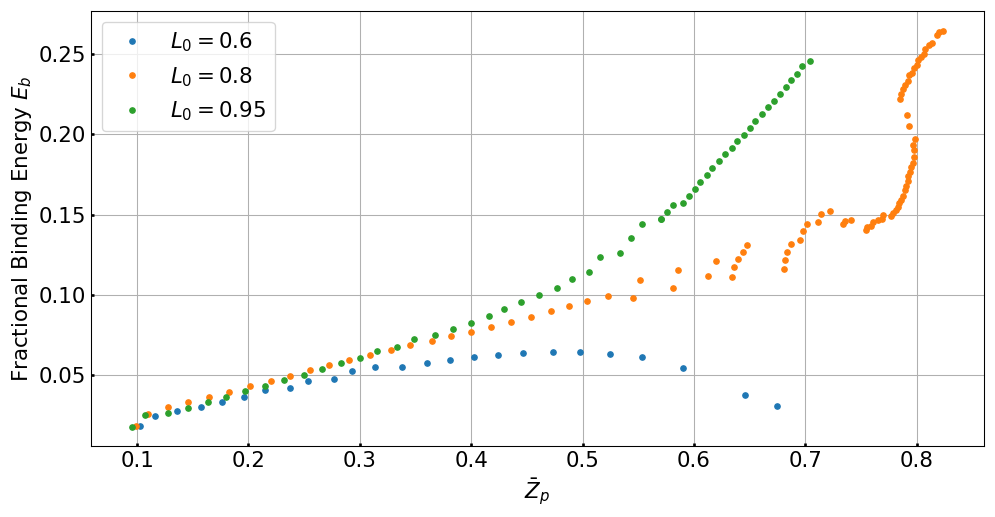}
        \caption{}
        \label{fig.SolutionSequences.EbvsZp}
    \end{subfigure}%
    ~
    \begin{subfigure}[t]{0.48\textwidth}
        \centering
        \includegraphics[height=4.5cm]{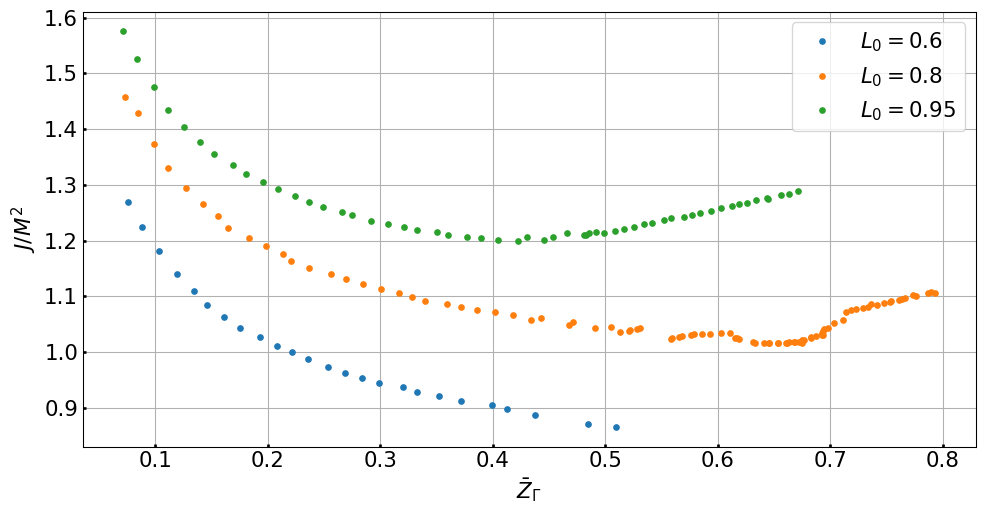}
        \caption{}
        \label{fig.SolutionSequences.JoverMsquared}
    \end{subfigure}%
    \vspace{0.2cm}
    \begin{subfigure}[t]{0.48\textwidth}
        \centering
	\includegraphics[height=4.5cm]{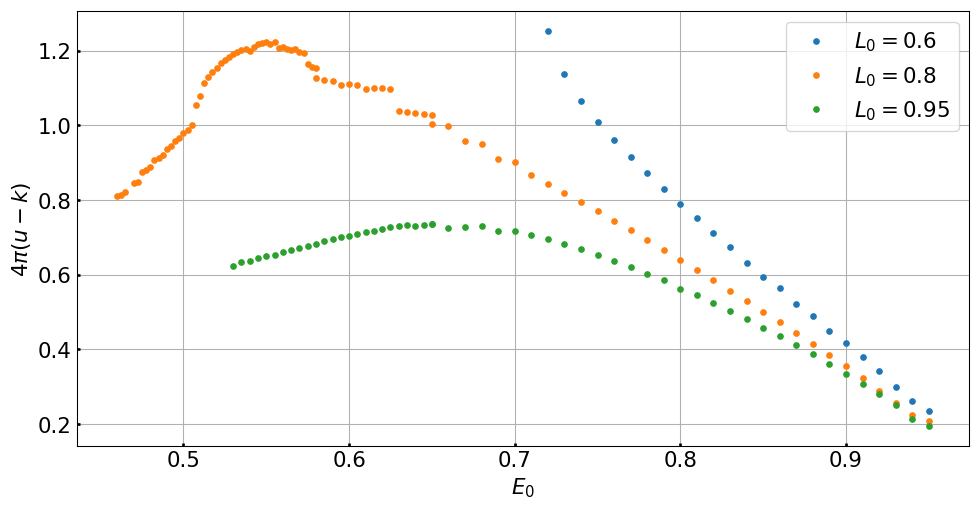}
        \caption{}
        \label{fig.SolutionSequences.HMVdeficit_vs_E0}
    \end{subfigure}
     ~
    \begin{subfigure}[t]{0.48\textwidth}
        \centering
	\includegraphics[height=4.5cm]{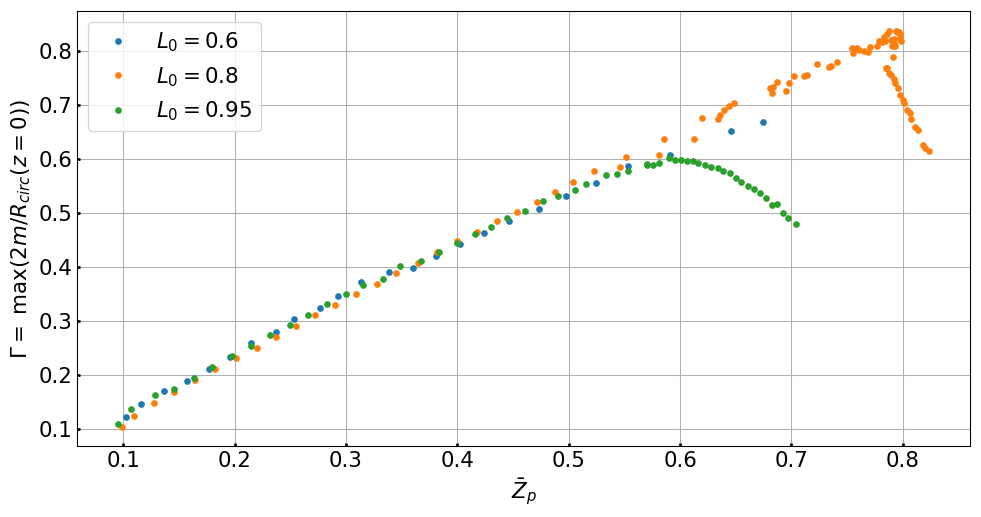}
        \caption{}
        \label{fig.SolutionSequences.MainSequence}
    \end{subfigure}%
    \caption{Solution characteristics for three different solution sequences. Definitions of the characteristics shown can be found in the text.
    Panel (a) shows the fractional binding energy versus the renormalized redshift at the peak in the energy density.
    Panel (b) shows the ratio of total angular momentum of total mass squared versus the renormalized redshift at the peak at the radius of maximum compactness.
    In panel (c) is plotted $4\pi$ times the linear energy density $u$ minus the azimuthal pressure $k$ versus $E_0$.
    Panel (d) displays the the compactness parameter versus the renormalized redshift at the peak in the energy density.}
\label{fig.SolutionSequences}
\end{figure}

To investigate the limiting behavior of solutions under the ansatz \Eqref{eq.ansatz}, we construct sequences of stationary solutions starting from an initial solution with diffuse density profile. For each sequence the $L_0$ parameter is fixed and the $E_0$ parameter is stepped down, where at each step the previous converged solution is provided as an initial guess to the solver. The behavior of various physical characteristics is investigated along the sequence, as well as compared with sequences of different $L_0$ parameter. Solution sequences for several values of $L_0$ were computed and in \Figref{fig.SolutionSequences} we illustrate the behavior of the $E_0, L_0$ parameter space with three sequences corresponding to $L_0 = 0.95, 0.8, 0.6$. Density profiles and accompanying ergoregions, if present, are shown in \Figref{fig.RHOErgo2D} and \Figref{fig.RHOErgo3D} for a selection of solutions on the $L_0 = 0.8$ sequence. Our studies indicate essentially two regimes, corresponding to low and high total angular momentum solutions. One can tune between these by varying the minimum particle angular momentum $L_0$. Evidence for two distinct relativistic limits, one in the high angular momentum regime, and one at the boundary of high and low angular momentum solutions, is presented below.

Solutions on low-angular momentum sequences (cf. $L_0 = 0.6$ sequence) become sub-critical in that $|\mathcal J|/ \mathcal M^2 < 1$ (see \Figref{fig.SolutionSequences.JoverMsquared}) as the support of the matter decreases. None of these solutions (including those along sequences with other $L_0$-values not shown) exhibit ergoregions. In all of the low angular momentum sequences the redshift $\bar Z$ (see \Eqref{eq.Redshift} and paragraph below) is monotonically increasing as $E_0$ is decreased. Along such sequences there is a maximum in the fractional binding energy $E_b$, as seen in \Figref{fig.SolutionSequences.EbvsZp}, and shortly after the maximum the residual in our numerical fixed-point iteration grows and the numerics break down. For the $L_0 = 0.6$ sequence this occurs after $\bar Z_p = 0.67$.
We speculate, consistent with a conjecture of Zel'dovich \cite{Zeldovich:gx}, that such solutions become unstable to black hole collapse. As mentioned in the introduction, there is evidence from studies in spherical symmetry that fixed-point iterations like the one employed here ``prefer'' stable solutions, and hence will avoid the unstable solutions past the maximum in the binding energy curve. Further work to test this instability numerically is in progress \cite{Ames:h5rQEYmq}.

Before discussing the two limits mentioned above we note the interesting property of the solution sequences illustrated in \Figref{fig.SolutionSequences.MainSequence}. Here it is shown that solutions can be classified as either lying on or off a \emph{main sequence}. Solutions on the main sequence correspond to those who's major axis (as measured by $\Rc$, cf. \Sectionref{sec.BHlimit} below) shrinks with increasing redshift, resulting in an increasing compactness parameter $\Gamma = 2m/R_{\mathrm circ}(z=0)$ (see text above \Eqref{eq.MassAspect} for a definition). At high redshift, such solutions appear to be approaching a black hole limit, as argued in \Sectionref{sec.BHlimit}. However, solution sequences with sufficiently large angular momentum (such as the $L_0 = 0.8$ and $L_0 = 0.95$ solution sequences shown in \Figref{fig.SolutionSequences.MainSequence}) eventually leave this sequence and tend towards a thin string limit. Properties of solutions at the extreme of this portion of the sequence are discussed in \Sectionref{sec.CosmicStringLimit}. In particular, the major axis grows (increasing $\Rc$, resulting in a decreasing compactness parameter $\Gamma$) while the minor axis of the configuration continues to decrease.
It is also interesting to observe that for solution sequences with ergoregions, those ergoregions first appear near $E_0 = 0.66$, independently of the $L_0$ parameter.

\subsection{A Quasi-stationary Transition to an Extreme Kerr Black Hole}
\label{sec.BHlimit}
Considering sequences with increasing $L_0$ values the angular momentum of the equilibrium solutions at a given redshift increases, as does the redshift at which the peak in the binding energy occurs. This suggests that there may be a $L_0$ for which one may approach a black hole solution ($\bar Z \to 1$) on the stable side of the binding energy curve and with sufficient angular momentum ($|\mathcal J|/\mathcal M^2 \ge 1$) such that the solution cannot collapse to a black hole. Experimentally we find this to occur for $L_0$ near $0.8$, for which (cf. \Figref{fig.SolutionSequences.JoverMsquared}) $|\mathcal J|/\mathcal M^2 \to 1$. It is natural to ask if the limiting member of this sequence is an extremal Kerr black hole, similar to the case of uniformly rotating fluid disks and rings mentioned in the introduction. \Figref{fig.SolutionSequences.JoverMsquared} shows that although $|\mathcal J|/\mathcal M^2$ approaches one near $\bar{Z}_{\Gamma} \approx 0.645$ it subsequently increases again. This is due to the solution sequence leaving the main sequence and transitioning to the thin string limit. This occurs near $E_0 = 0.545$. In investigating the black hole limit we restrict our attention to solutions prior to this on the sequence.

\begin{figure}[t!]
    \centering
    \begin{subfigure}[t]{0.48\textwidth}
        \centering
        \includegraphics[height=4.0cm]{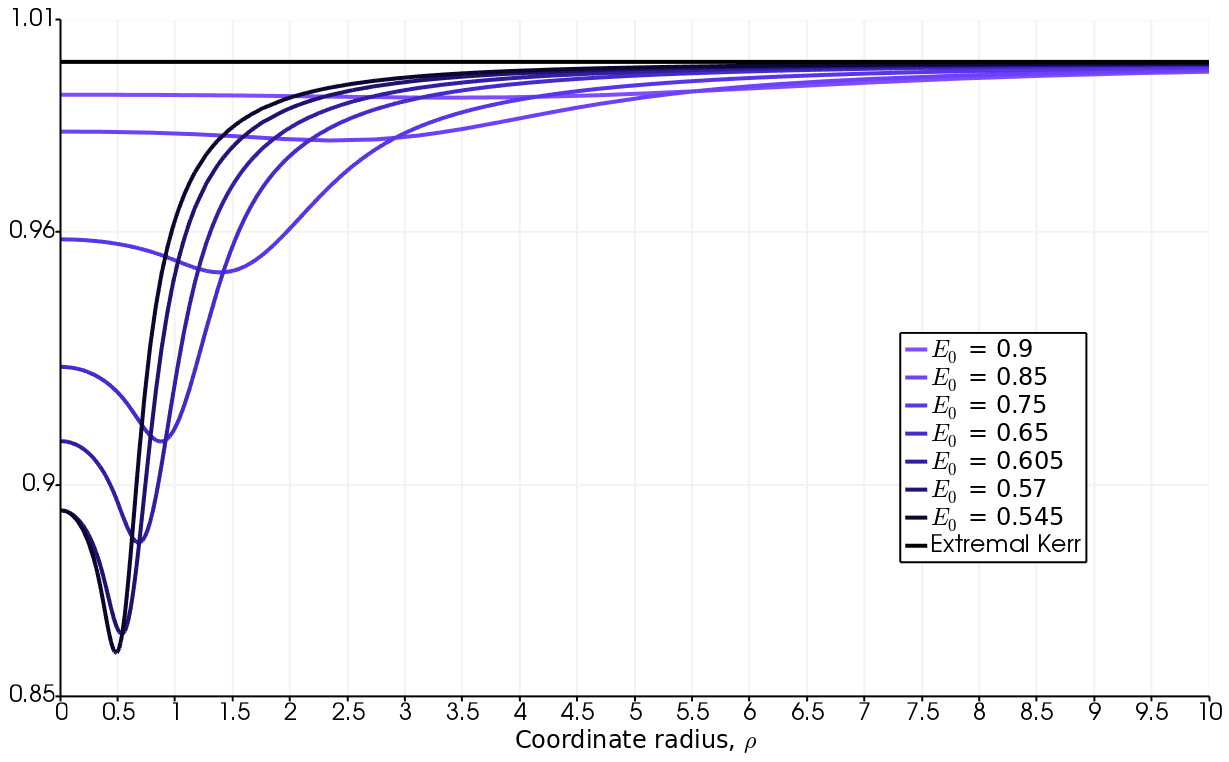}
        \caption{}
        \label{fig.MetricFieldCrossSections.BH.BB}
    \end{subfigure}%
    ~
    \begin{subfigure}[t]{0.48\textwidth}
        \centering
        \includegraphics[height=4.0cm]{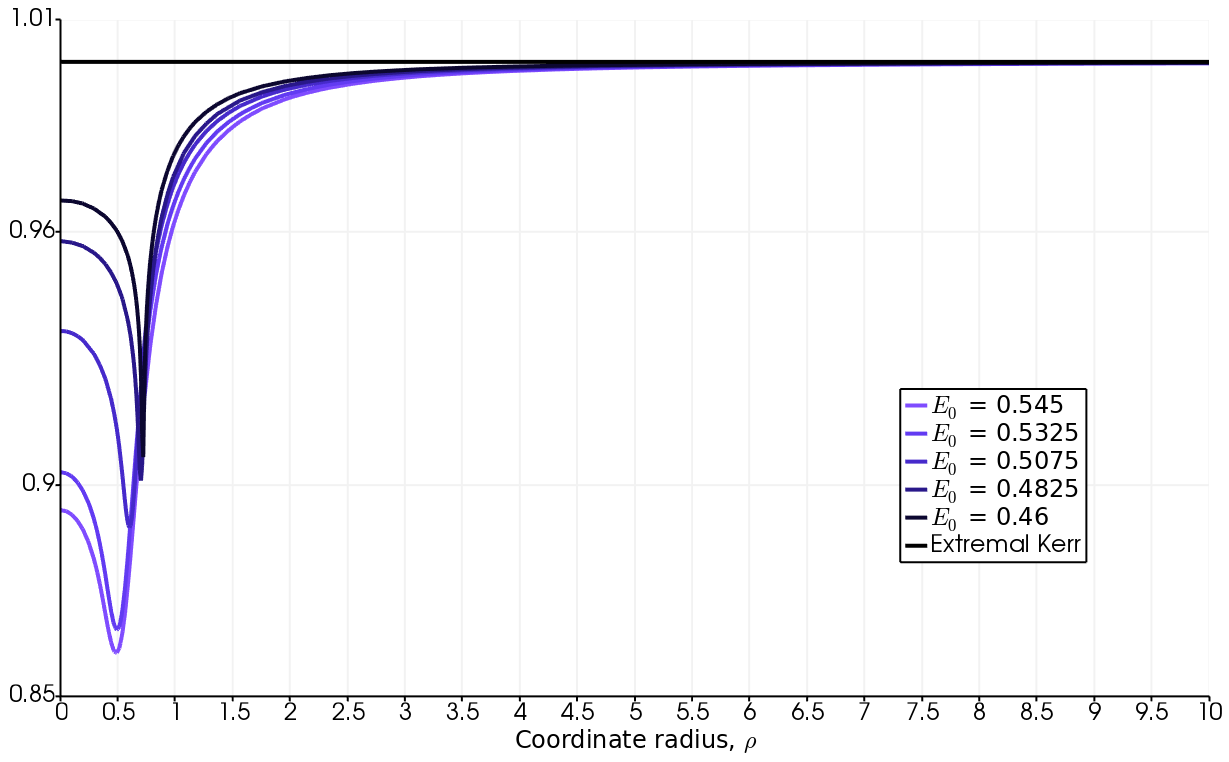}
        \caption{}
        \label{fig.MetricFieldCrossSections.String.BB}
    \end{subfigure}%
    \vspace{0.2cm}
    \begin{subfigure}[t]{0.48\textwidth}
        \centering
        \includegraphics[height=4.0cm]{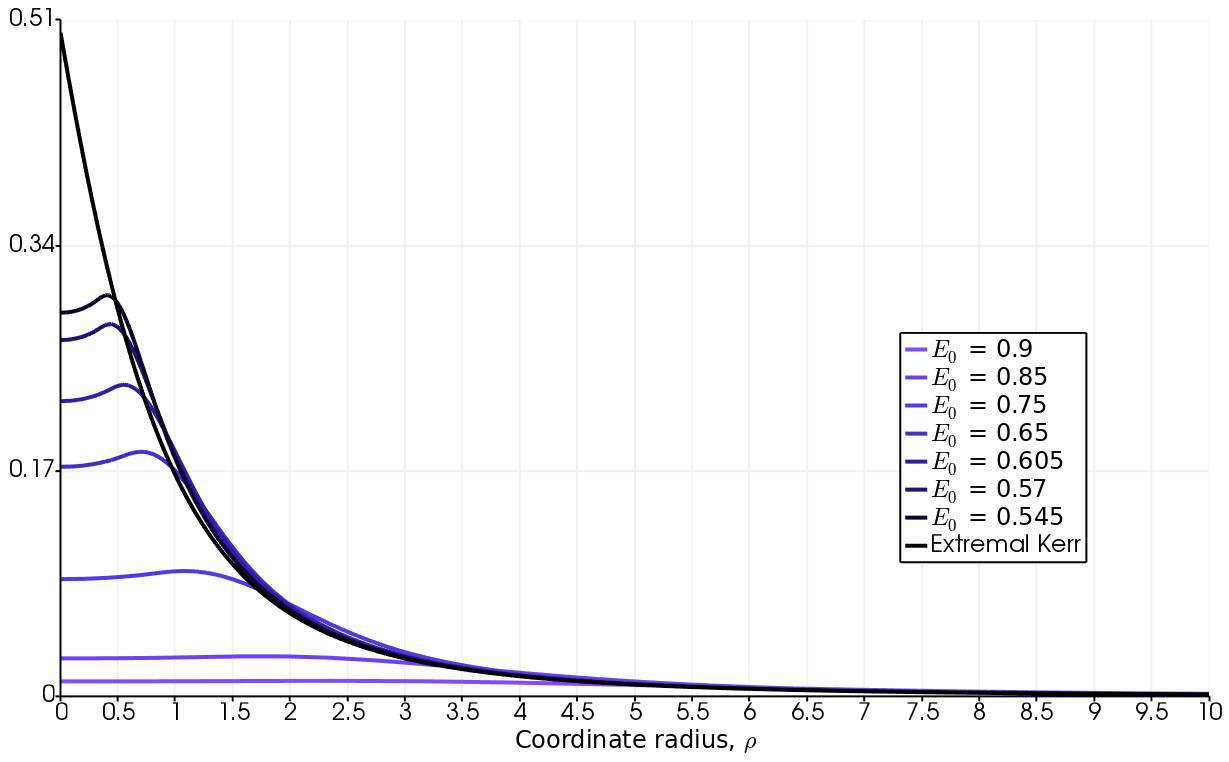}
        \caption{}
        \label{fig.MetricFieldCrossSections.BH.WW}
    \end{subfigure}
     ~
    \begin{subfigure}[t]{0.48\textwidth}
        \centering
        \includegraphics[height=4.0cm]{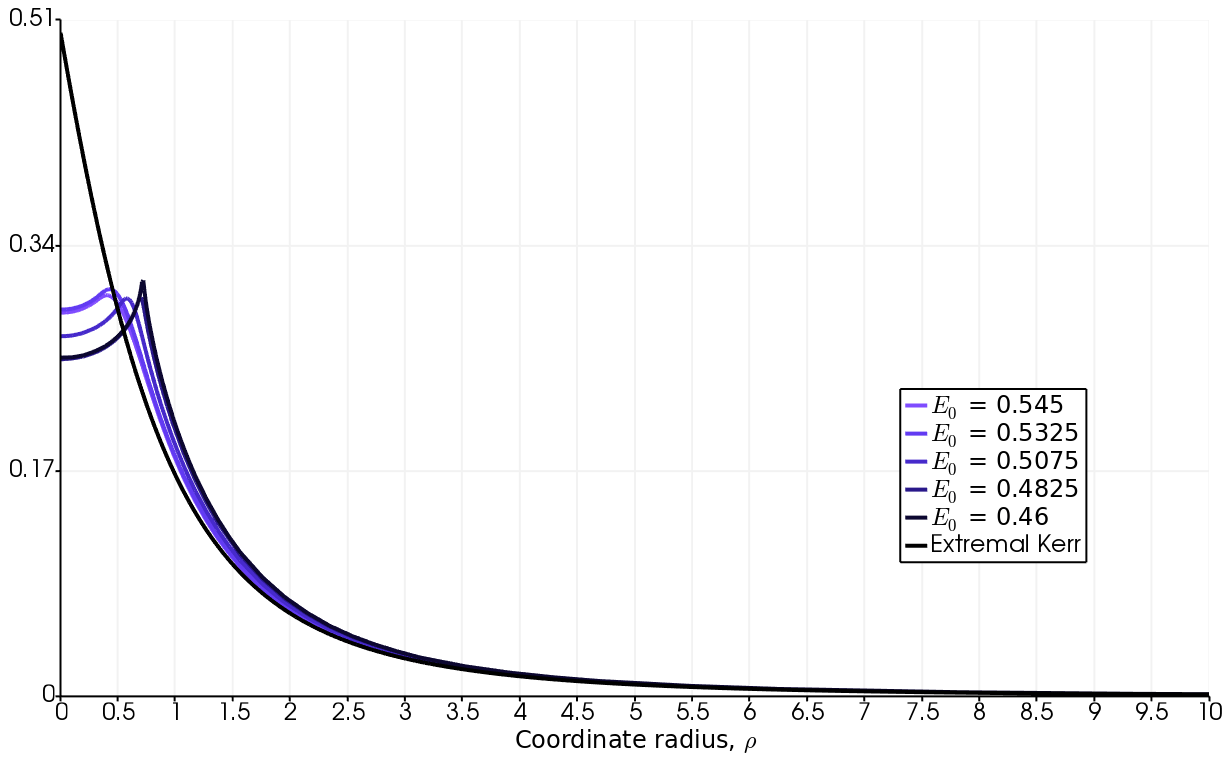}
        \caption{}
        \label{fig.MetricFieldCrossSections.String.WW}
    \end{subfigure}%
        \vspace{0.2cm}
    \begin{subfigure}[t]{0.48\textwidth}
        \centering
        \includegraphics[height=4.0cm]{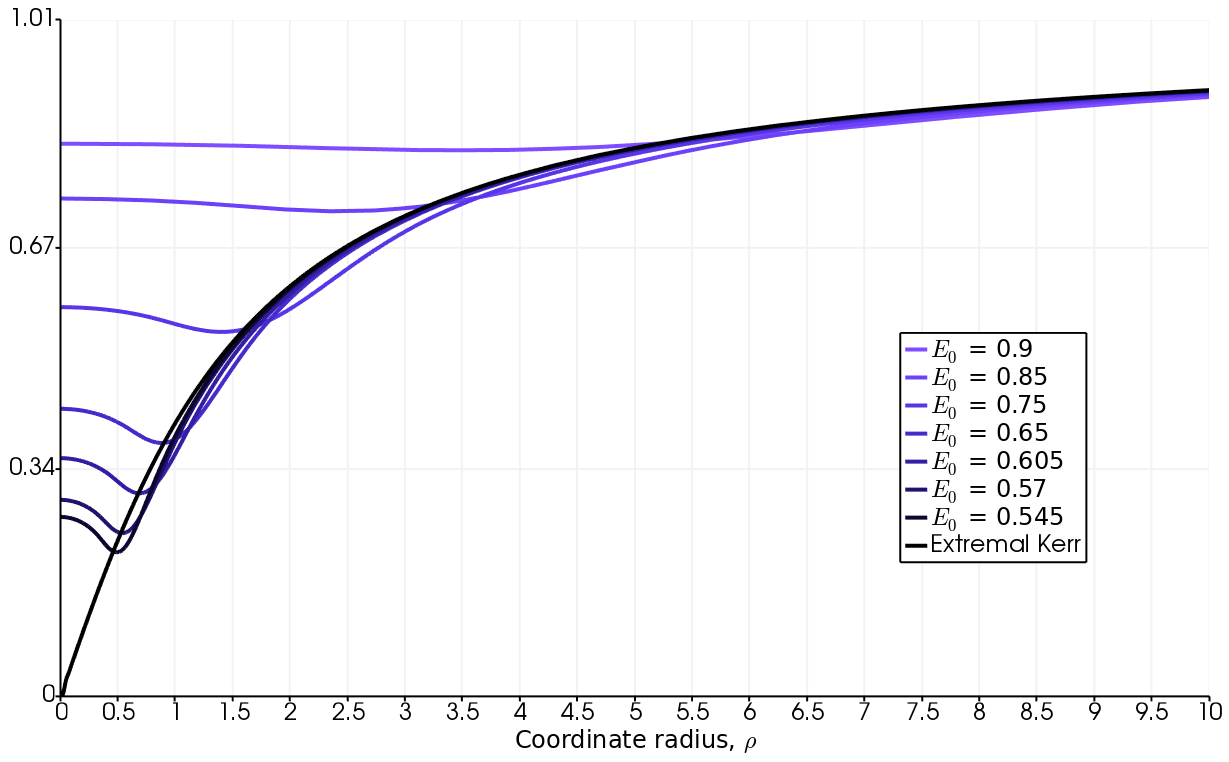}
        \caption{}
        \label{fig.MetricFieldCrossSections.BH.NU}
    \end{subfigure}
     ~
    \begin{subfigure}[t]{0.48\textwidth}
        \centering
        \includegraphics[height=4.0cm]{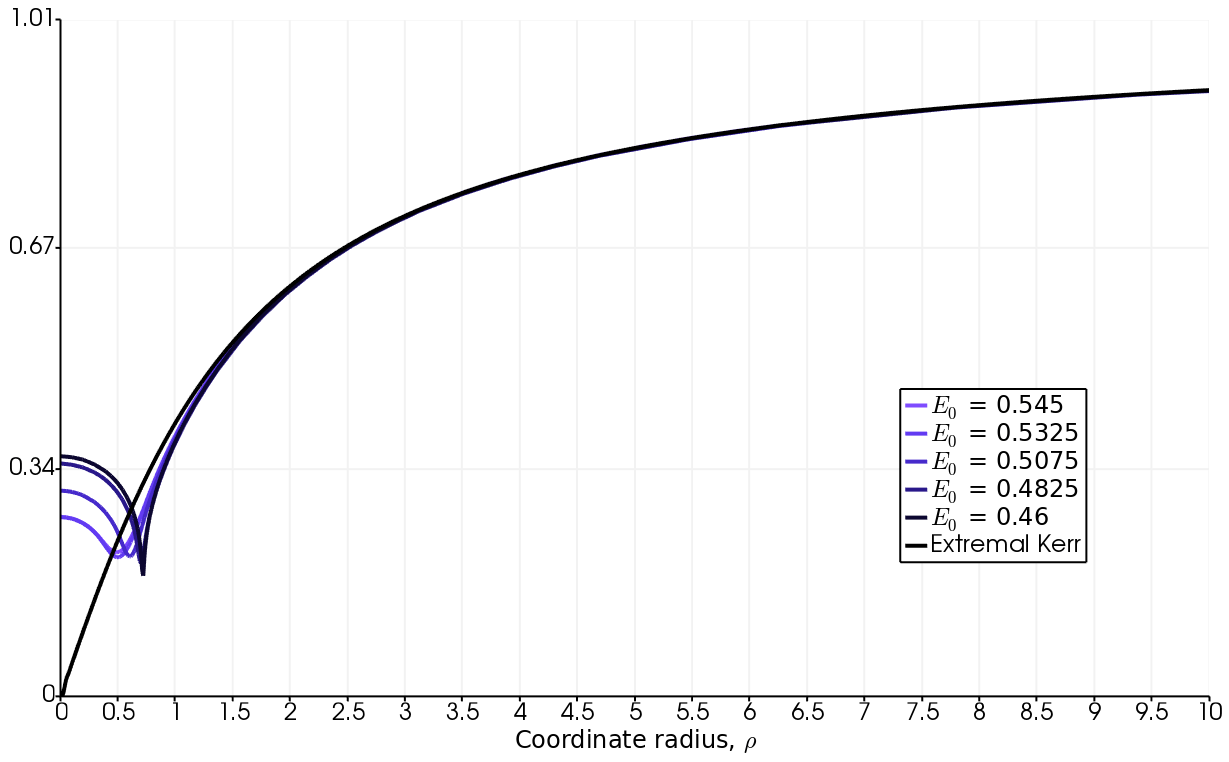}
        \caption{}
                \label{fig.MetricFieldCrossSections.String.NU}
    \end{subfigure}%
\caption{Cross-sections in the reflection plane ($z=0$) of the metric fields for a selection of solutions, and comparison to the extreme Kerr solution (black). Panels (a) and (b) show the $B$ field, panels (c) and (d) the $\omega$ field, and $e^\nu$ is shown in panels (e) and (f).}
\label{fig.MetricFieldCrossSections}
\end{figure}

The limiting behavior of the metric fields are investigated and compared with that of the extremal black hole solutions in \Figref{fig.MetricFieldCrossSections}. With the coordinates and metric parametrization used here, the extreme Kerr solution of mass $\mathcal M$ has a metric of the form \Eqref{eq:Metric} with \cite{Bardeen:1971cz,Ansorg:2008jr}
\begin{align}
\label{eq.NU.extremeKerr}
\nu_{EK} & =
\frac 12 \ln \left( \frac{r^2 ( \mathcal M + r)^2 + \mathcal M^2 z^2}{(\mathcal M^2 + (\mathcal M + r)^2 )^2 - \mathcal M^2 \rho^2}\right), \\
\label{eq.BB.extremeKerr}
B_{EK} & = 1, \\
\mu_{EK} & = \frac 12 \ln \left( r^{-2} (\mathcal M + r)^2 + \mathcal M^2 z^2 r^{-4} \right), \\
\omega_{EK} & = \frac{2 \mathcal M^2( \mathcal M + r)}{(\mathcal M^2 + (\mathcal M + r)^2 )^2 - \mathcal M^2 \rho^2},
\end{align}
where $r^2 = \rho^2 + z^2$. In these coordinates the horizon is at $\rho = z = 0$, and the exterior vacuum spacetime is coordinatized on the domain $\rho \in (0, \infty), z \in (-\infty, \infty)$. The $r \to 0$ limit implies $\nu_{EK} \to -\infty$, $\mu_{EK} \to \infty$, and $\omega_{EK} \to 1/(2\mathcal M)$. The fields $B$ (\Figref{fig.MetricFieldCrossSections.BH.BB}), and $\omega$ (\Figref{fig.MetricFieldCrossSections.BH.WW}), are regular at the horizon and thus amenable to direct comparison. For the $\nu$ field an appropriate quantity to compare is the lapse $e^\nu$ (\Figref{fig.MetricFieldCrossSections.BH.NU}), which vanishes at the origin and approaches one asymptotically. A construction of the $\mu$ field which is regular at the origin also grows rapidly, making the visual comparison less useful. Such a plot has therefore been omitted. The plots in \Figref{fig.MetricFieldCrossSections} show cross sections along the reflection plane ($z=0$) of the metric fields for several solutions. Solutions on the main sequence exhibit steady approach to the extremal Kerr solution.
We note that similar plots (cf. \Figref{fig.MetricFieldCrossSections} panels (b, d, f)) show that after $E_0 \approx 0.545$ the fields cease this approach to the extremal Kerr solution and appear to converge to a different solution.

In spherical symmetry a black hole forms if the mass $\mathcal M$ becomes confined within a Schwarzschild radius of $\mathcal R = 2 \mathcal M$. There is no such well-defined criteria in axisymmetry, although the Hoop conjecture \cite{Thorne:1972th} puts forward that a non-spherical body with mass $2\mathcal M$ exceeding its corresponding Schwarzschild radius $\mathcal R$ and which is confined within a ball of radius $\mathcal R$ is a necessary and sufficient for the formation of a black hole. In our setting, a natural measure of the radius is the length of the axisymmetric Killing vector field which we denote $\Rc := \rho B e^{-\nu}$. This quantity provides a natural length scale for the solution, in particular when restricted to the reflection plane ($z=0$) and evaluated near the boundary of the matter. For Vlasov matter, which typically has an extended atmosphere, it is useful to take the radius at which the integrated mass density over the radius is maximum. We define the compactness parameter $\Gamma := \max_{\rho \in (0, \infty)} 2 m(\rho)/\bar{R}_{\mathrm{circ}}(\rho)$, where $\bar{R}_{\mathrm{circ}} := (\Rc )|_{z=0}$ and where
\begin{equation}
\label{eq.MassAspect}
m(\rho) := 2\pi \int_{z = -\infty}^\infty \int_{\tilde\rho = 0}^\rho B\left(\Phi_{00} + \Phi_{11} + \Phi_{33}(1 - (\tilde\rho B)^2\omega^2 e^{-4\nu})\right)\tilde\rho \, \md \tilde\rho \dz.
\end{equation}
Note that $m(\rho) = \mathcal M$ when $\rho$ exceeds the matter support. For the regular solutions we construct $\Gamma \in (0, 1)$.
A couple of remarks on $\Gamma$:
(\emph{i}) Using the expressions \Eqref{eq.NU.extremeKerr} and \Eqref{eq.BB.extremeKerr} above for $B$ and $\nu$ in the extreme Kerr limit, where the black hole is represented by a point at the origin, it can be shown that $\bar{R}_{\mathrm{circ}}(\rho)$ is an increasing function of $\rho$ and that $\lim_{\rho \to 0} \bar{R}_{\mathrm{circ}}(\rho) = 2\mathcal M$. Thus, for an extremal Kerr black hole $\Gamma = 1$.
(\emph{ii}) Using $\Rc$ as a radius in the hoop conjecture, one obtains that a black hole forms if and only if $2 \mathcal M/\Rc \ge 1$.
These remarks suggest that the $\Gamma \to 1$ is a black hole limit. As shown in \Figref{fig.SolutionSequences.MainSequence}, for the $L_0 = 0.8$ sequence $\Gamma$ increases to approximately $0.82$ at $E_0 = 0.545$. Our solution sequence then bends towards the thin string limit, for which $\Gamma$ decreases due to an increasing support of the matter.

Another characterization of a relativistic solution geometry is the redshift. We characterize the redshift using the ZAMO class of observers \cite{Bardeen:1972ip} for which the redshift of a photon emitted at $(\rho,z)$ and observed at spatial infinity is given by $Z(\rho, z) = e^{-\nu(\rho, z)} -1$. For solutions approaching a black hole, the field $\nu$ (which is negative) becomes unbounded, and it is useful to instead use the rescaled quantity
\begin{equation}
\label{eq.Redshift}
 \bar Z := \frac{Z}{1 + Z} = 1- e^{\nu},
\end{equation}
which in the black hole limit approaches one. We give values for this quantity at the peak density of the body, denoted $\bar Z_p$, and at the radius at which $\Gamma$ achieves its maximum, denoted $\bar Z_\Gamma$. Measured at the peak in the density, $\bar Z_p$ increases to $\bar Z_p \approx 0.78$ at $E_0 = 0.545$. The fact that the $L_0=0.8$ sequence approaches both $\Gamma = 1$ and $\bar Z_p = 1$ (cf. \Figref{fig.SolutionSequences.MainSequence}) as $\mathcal J/\mathcal M^2 \to 1$ suggests that such a family of solutions exhibits a quasistationary transition to an extreme black hole. It is likely that stationary solutions close to the black hole are unstable, and the fact that we are unable to obtain more relativistic solutions is, we speculate, due to the preference of our numerical scheme for stable solutions.

\subsection{Self-Consistent Cosmic String Models}
\label{sec.CosmicStringLimit}
Families of solutions in the super-critical regime, $|\mathcal J|/\mathcal M^2 > 1$, eventually tend towards a limit which is distinct from the black hole limit. Due to the thin ring-like nature of these solutions, as well as properties of the spacetime discussed below, we refer to this as the \emph{string limit}. Such sequences are illustrated in \Figref{fig.SolutionSequences} by the $L_0 =0.95$ solution sequence. The near-critical $L_0 = 0.8$ solution sequence also tends towards a string limit late in the solution sequence. In fact, the string limit appears attractive to the code in the sense that for parameters $L_0 > 0.8$ the string limit is approached increasingly earlier in the sequence.

\begin{figure}[th!]
\begin{center}
\includegraphics[width=.5\linewidth]{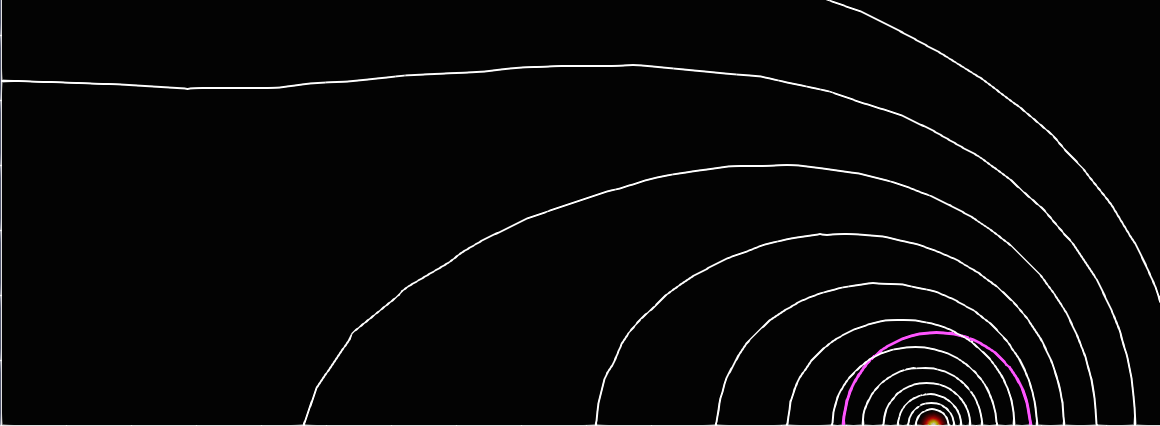}
\caption{Near circular contours of the $\mu$ field in the region near the matter for the $L_0 = 0.8, E_0 = 0.46$ solution. A $\sigma = 3$ contour of the toroidal radius is included for reference (magenta curve).}
\label{fig.MUContoursRHO}
\end{center}
\end{figure}

After branching off from the main sequence (cf. \Figref{fig.SolutionSequences.MainSequence}) an $E_0$-parameterized sequence of solutions becomes increasingly thin and the radius of the peak density increases slightly, as can be seen in \Figref{fig.RHOErgo3D} and \Figref{fig.RHOErgo2D} panels (h)--(l). 
The geometry of extreme solutions in this limit has a near-field regime which is locally rotationally symmetric about the ring and conical. This regime is small compared to the ring radius, as illustrated by the contour plot \Figref{fig.MUContoursRHO}. Far away, the fields are, as dictated by the boundary conditions, asymptotically flat. Additionally, a computation of the Kretschmann scalar indicates that it vanishes a distance of order the ring radius from the ring. One notes from \Figref{fig.MetricFieldCrossSections.String.BB}, \Figref{fig.MetricFieldCrossSections.String.NU}, and \Figref{fig.MetricFieldCrossSections.String.WW} that along the string sequence the fields move away from the black hole limit and apparently converge to a distinct limiting configuration.

The primary feature which characterizes the conical geometry is the deficit angle as one circumnavigates the matter torus in the meridional plane, which we denote by $\Delta \eta$. To measure this angle and facilitate comparison with previous studies on circular cosmic strings in the literature \cite{Frolov:1989ix,Hughes:1993ur,McManus:1993fp}, we write the relevant part of our metric in toroidal coordinates $(\sigma, \psi) \in (0, \infty) \times [-\pi, \pi)$ defined by
\[ \rho = a N^{-2} \sinh(\sigma), \quad z = a N^{-2} \sin(\psi),\]
where
\[ N^2 = \cosh(\sigma) - \cos(\psi) \]
and $a = \rho_{\text{peak}}$ is the radial coordinate of the peak density. In these coordinates the peak in the density is obtained by the limit $\sigma \to \infty$. The metric in the meridional plane takes the form
\[e^{2 \tilde\mu} a^2 N^{-4} (\md \sigma^2 + \md \psi^2), \]
where $\tilde\mu(\sigma, \psi) = (\mu \circ y) (\sigma, \psi)$ and $y$ is the coordinate map $(\sigma, \psi) \mapsto (\rho, z)$.
The deficit angle can be expressed in terms of the ratio of the proper $\psi$-arclength to proper $\sigma$-radius
\begin{equation}
\label{eq.deficit_angle}
\Delta \eta |_{\sigma = \sigma_0, \psi = \psi_0} = 2\pi - \frac{\int_{-\pi}^\pi \left(aN^{-2}(\sigma, \psi)e^{\tilde\mu(\sigma, \psi)}\right)|_{\sigma = \sigma_0} \md \psi} {\int_{\sigma_0}^\infty \left(aN^{-2}(\sigma, \psi)e^{\tilde\mu(\sigma, \psi)}\right)|_{\psi = \psi_0} \md \sigma}.
\end{equation}
This quantity is based at $(\sigma_0, \psi_0)$ in the sense that one must choose values at which to evaluate the integrals. The results of this calculation for a selection of solutions in the $L_0 = 0.8$ solution sequence is shown in \Figref{fig.DeficitAngle}, where in each panel we plot $\Delta \eta$ versus $\sigma_0$ for $\psi_0 = (0, \pi/3, 2\pi/3, \pi)$, corresponding to the different color traces. We say that the deficit angle is well-defined where these traces agree, and not defined otherwise, presumably due to a deviation of the geometry from locally rotationally symmetric about the matter.

The vertical line in each panel is an estimate of boundary of the matter $\sigma_{\text{supp}}$, which is obtained by
\begin{equation}
\label{eq.sigma_supp}
\sigma_{\text{supp}} :=
\min_{\tilde \rho \in \{ \rho_{\text{inner}}, \, \rho_{\text{outer}} \}}
\left\{ \log \left(  (\rho_{\text{peak}} + \tilde\rho)^2 / (\rho_{\text{peak}} - \tilde\rho)^2 \right) \right\}.
\end{equation}
Recall that a smaller $\sigma$ corresponds to a further distance from the center of the ring. For reference, the $\sigma = 3$ contour is included (magenta line) in the contour plot \Figref{fig.MUContoursRHO}.
The plots in \Figref{fig.DeficitAngle} have been cut-off at $\sigma = 12$ in order to show more detail at low $\sigma$. However, $\Delta \eta$ also remains zero to $\sigma$ as large as our numerical resolution allows.
The horizontal line in each plot is discussed below.

\newcommand{\tw}{0.32\textwidth}
\begin{figure}[t!]
    \centering
    \begin{subfigure}[t]{\tw}
        \centering
        \includegraphics[height=3cm]{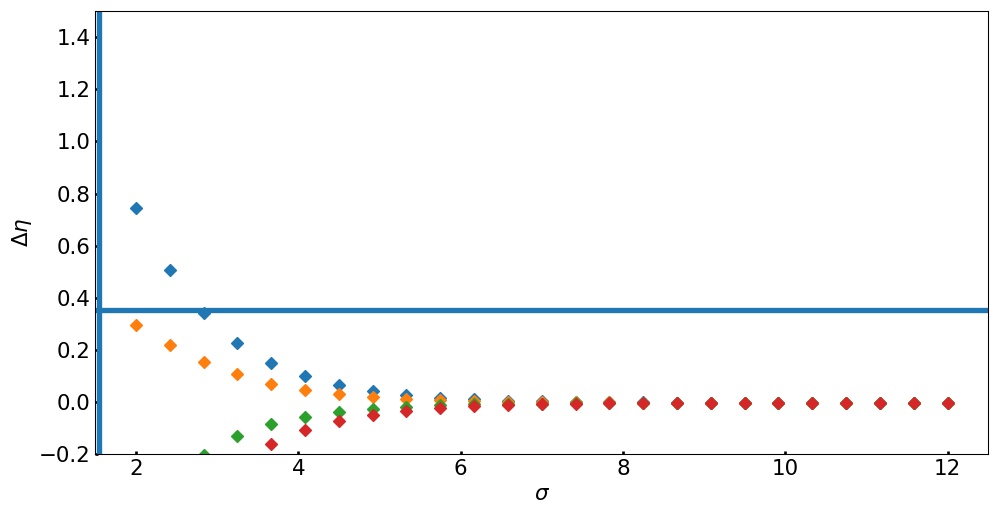}
        \caption{$E_0 = 0.9$}
        \label{fig.DeficitAngle.05}
    \end{subfigure}%
    ~
    \begin{subfigure}[t]{\tw}
        \centering
        \includegraphics[height=3cm]{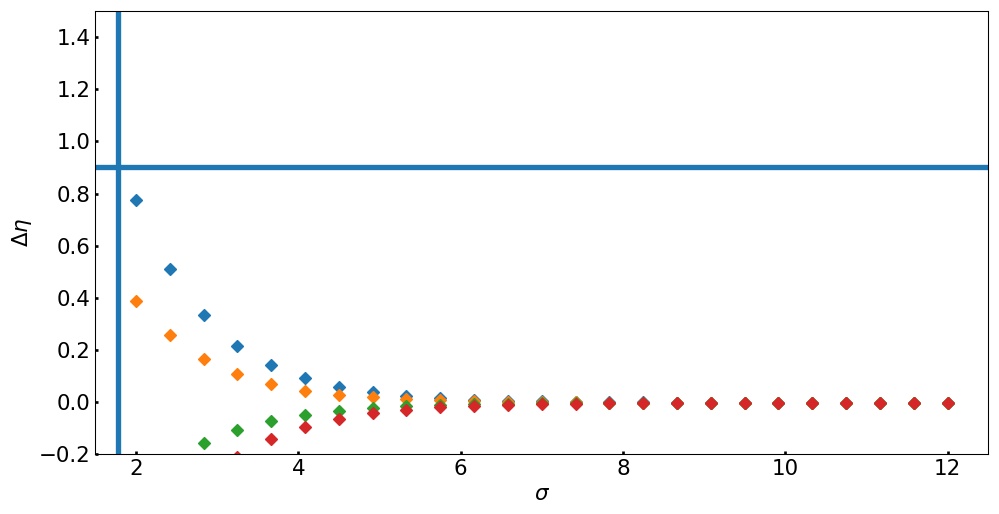}
        \caption{$E_0 = 0.7$}
    \end{subfigure}%
    ~
    \begin{subfigure}[t]{\tw}
        \centering
        \includegraphics[height=3cm]{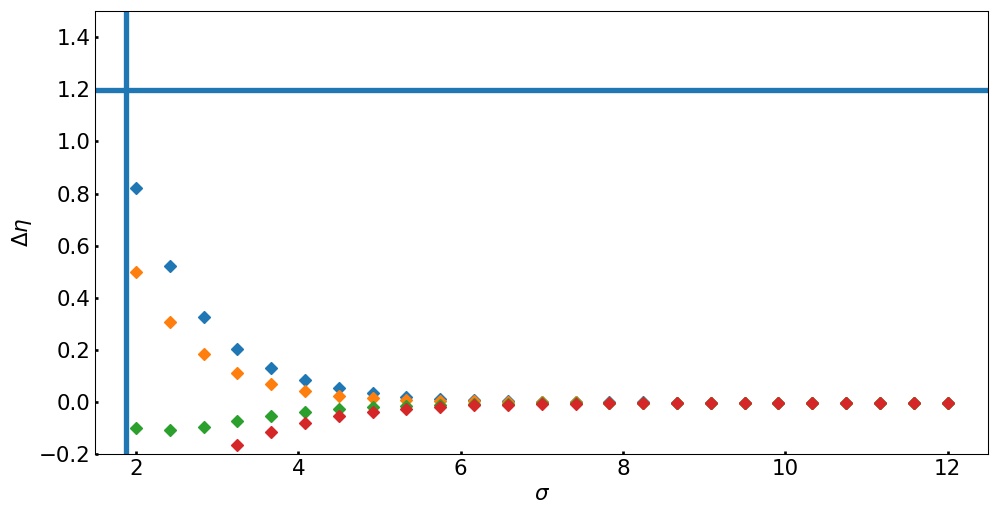}
        \caption{$E_0 = 0.57$}
    \end{subfigure}

    \vspace{0.2cm}

    \begin{subfigure}[t]{\tw}
        \centering
        \includegraphics[height=3cm]{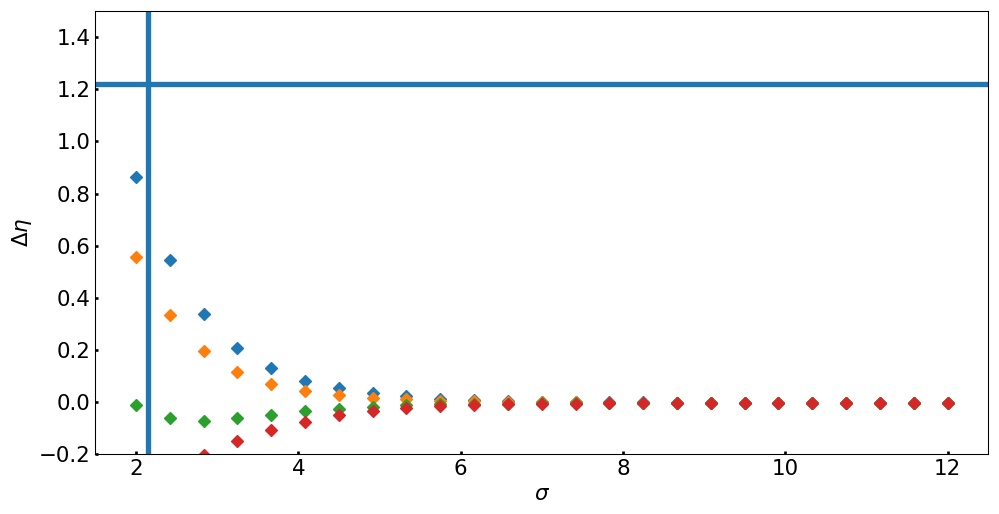}
        \caption{$E_0 = 0.545$}
    \end{subfigure}%
    ~
    \begin{subfigure}[t]{\tw}
        \centering
        \includegraphics[height=3cm]{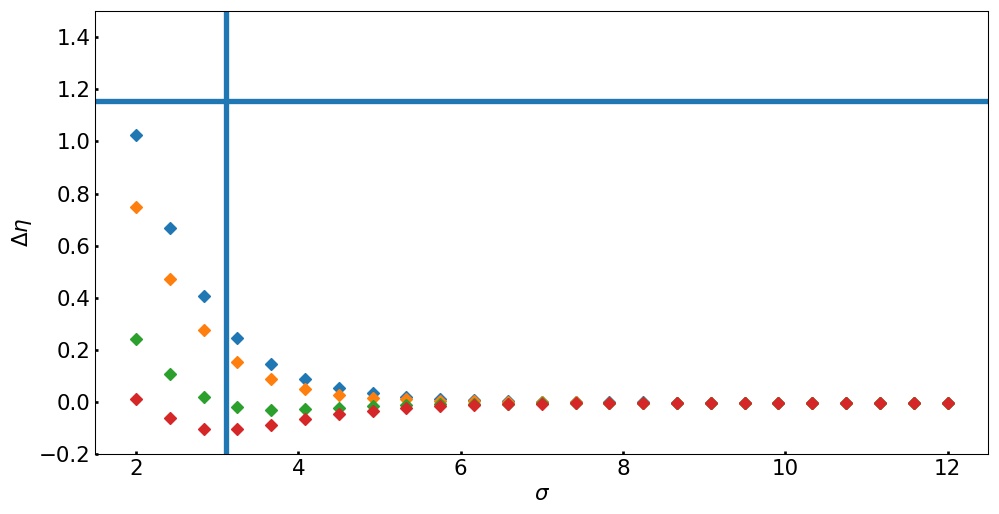}
        \caption{$E_0 = 0.52$}
    \end{subfigure}%
    ~
    \begin{subfigure}[t]{\tw}
        \centering
        \includegraphics[height=3cm]{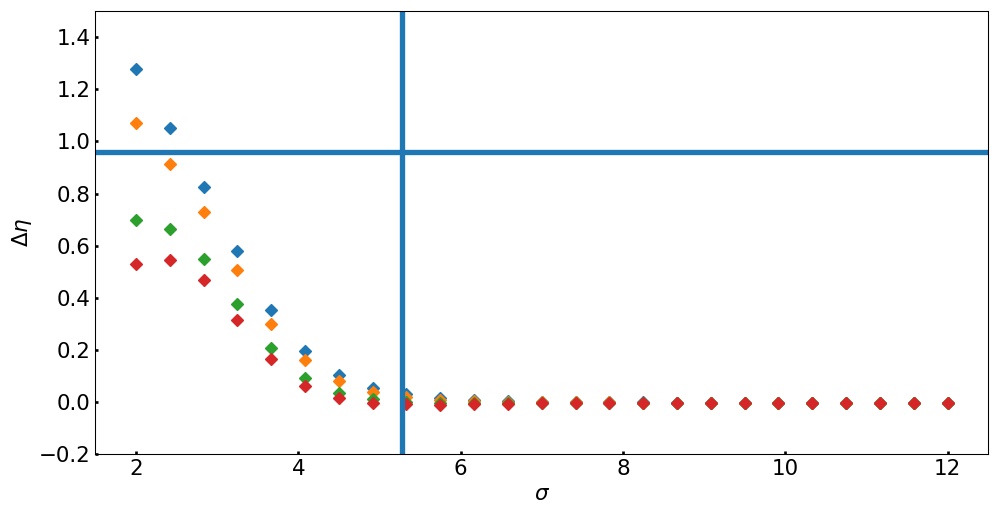}
        \caption{$E_0 = 0.495$}
    \end{subfigure}

    \vspace{0.2cm}

    \begin{subfigure}[t]{\tw}
        \centering
        \includegraphics[height=3cm]{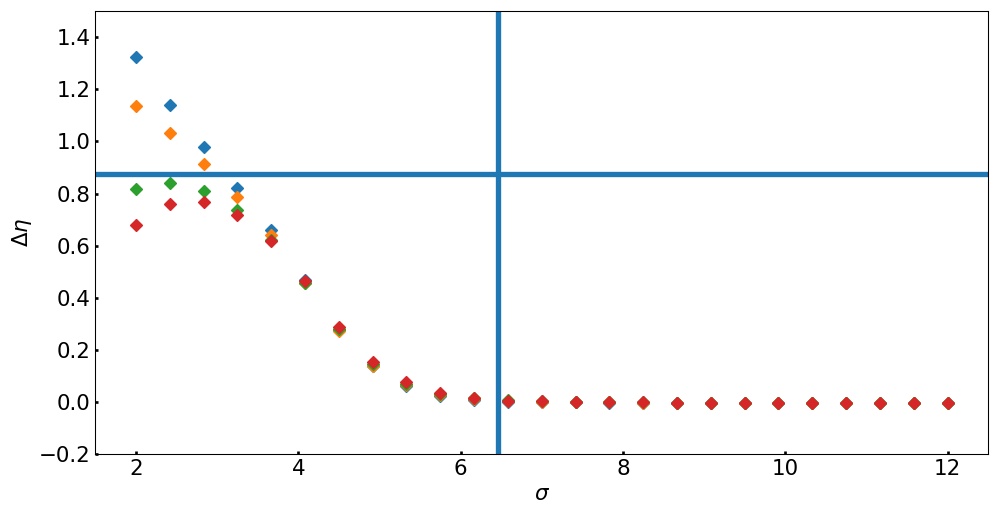}
        \caption{$E_0 = 0.475$}
    \end{subfigure}%
    ~
    \begin{subfigure}[t]{\tw}
        \centering
        \includegraphics[height=3cm]{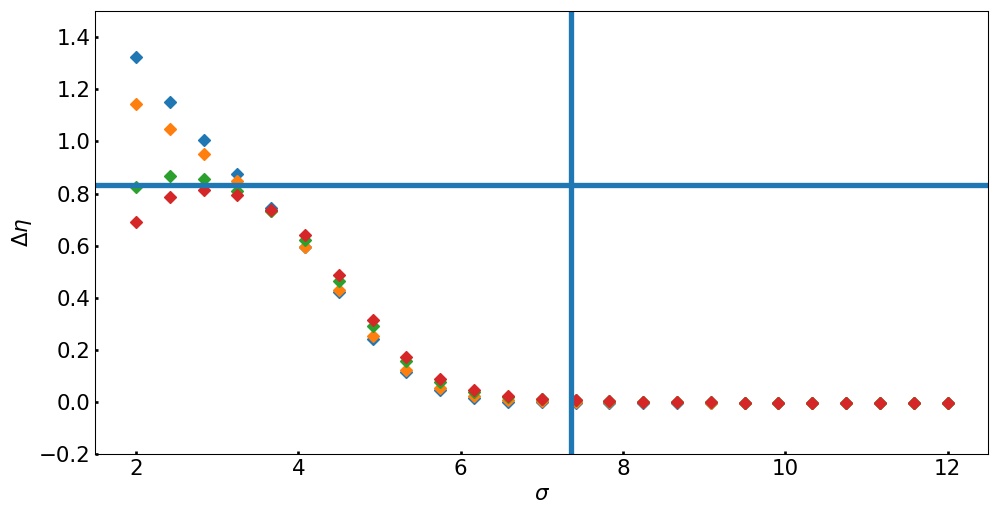}
        \caption{$E_0 = 0.4675$}
    \end{subfigure}%
    ~
    \begin{subfigure}[t]{\tw}
        \centering
        \includegraphics[height=3cm]{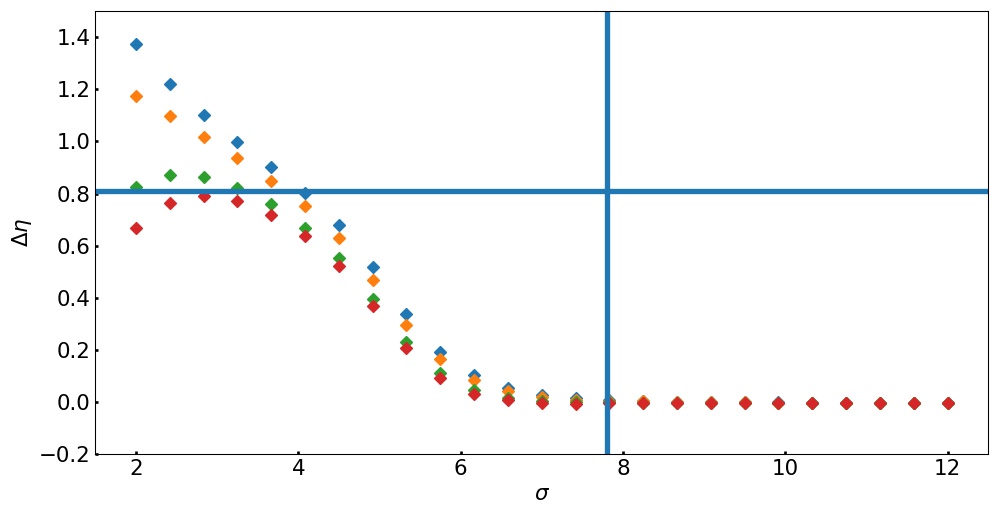}
        \caption{$E_0 = 0.46$}
        \label{fig.DeficitAngle.94}
    \end{subfigure}
    \caption{Deficit angle as computed with \Eqref{eq.deficit_angle} versus $\sigma_0$ for $\psi_0 = (0, \pi/3, 2\pi/3, \pi)$ for a selection of solutions on the $L_0 = 0.8$ solution sequence. In each panel the vertical line indicates the support of the matter computed by \Eqref{eq.sigma_supp}, and the horizontal line is an estimate of the deficit angle based on linearized theory for Dirac sources --see \Eqref{eq.LinearDiracDeficitEstimate} and also \Figref{fig.SolutionSequences.HMVdeficit_vs_E0}. }
    \label{fig.DeficitAngle}
\end{figure}

Inspection of \Figref{fig.DeficitAngle} shows that for solutions on the main sequence (those with $E_0 > 0.545$) the deficit angle is zero where it is defined, and becomes undefined even within the support of the matter due to a lack of symmetry for such solutions. As the solution sequence transitions to the string limit, $\Delta \eta$ begins to increase outside of the support of the matter. This angle grows with the distance from the matter to a maximum value before the geometry transitions to a regime which is no longer rotationally symmetric about the matter and the deficit angle is no longer defined. This transition occurs near $\sigma \approx 3$, from which one obtains that the radius of the conical region (centered on the peak density) is roughly a tenth of the coordinate radius of the ring (cf. \Figref{fig.MUContoursRHO}). This defines the near-field conical regime.

The near and far-field geometry exhibited by the solutions close to the string limit are consistent with investigations of circular cosmic string spacetimes obtained by Frolov et al. \cite{Frolov:1989ix}. There the authors impose a conical singularity in the metric and study the relationship between the deficit angle and various measures of the mass of the spacetime. 
As pointed out by Garfinkle and coauthors \cite{Garfinkle:1985ui,Futamase:1988iu,Garfinkle:1989gw}, in general relativity the gravitational field should be found by solving the coupled Einstein-matter system for an appropriate matter model rather than through prescribing a fixed energy-momentum tensor. They find however, in the case of certain Einstein-scalar-gauge field models for infinitely long straight cosmic strings, that the deficit angle can be approximated by the result obtained through a prescribed Dirac energy-momentum and linearized gravity originally derived by Vilenkin \cite{Vilenkin:1981ke}. 
In the case of circular cosmic strings Hughes et al. \cite{Hughes:1993ur} and McManis and Vandyck \cite{McManus:1993fp} have adapted the approach of Vilenkin with a prescribed distributional energy momentum tensor to circular cosmic strings. They specify a Dirac source at $\rho = a, z=0$ with a linear energy density $u = -T_t{}^t$ and linear azimuthal pressure $k = T_\varphi{}^\varphi$. Using the linearized Einstein equations it is found---both without \cite{Hughes:1993ur} and with \cite{McManus:1993fp} rotation---that the deficit angle $\Delta \eta$ can be approximated by $4\pi(u - k)$ (setting $G=1$). 

Given the results just stated it is of interest to compare the deficit angle for the solutions constructed in this paper with the results in \cite{Hughes:1993ur,McManus:1993fp} obtained from a prescribed Dirac source. Since the solutions presented in this paper have regular (non-distributional) energy momentum, in order to compare with these results we compute the corresponding quantities $u$ and $k$ by integrating the components of the energy momentum tensor over the meridional plane. Using the metric \Eqref{eq:Metric} and the integration measure $e^{2\mu} \md \rho \md z$ on the meridional plane we find
\begin{align}
u &= \int_{\mathbb R^2}  e^{2\mu - 2\nu} \left( T_{tt} + \omega T_{t\varphi} \right) \md \rho \md z, \\
k &= \int_{\mathbb R^2}  \left(
	(\rho B)^{-2} e^{2\mu + 2\nu} (1 - (\rho B)^2 \omega^2 e^{-4 \nu}) T_{\varphi \varphi}
	- \omega e^{2\mu - 2\nu} T_{t\varphi}
	\right) \md \rho \md z.
\end{align}
Subtracting and writing in terms of the $\Phi$-quantities (cf. \Eqref{eq.PHIs}) we obtain
\begin{equation}
\label{eq.LinearDiracDeficitEstimate}
4\pi(u-k)  = 4\pi \int_{\mathbb R^2} \left(
	\Phi_{00} + 2 \omega e^{-4\nu} \Phi_{03} - (1 - (\rho B)^2 \omega^2 e^{-4\nu})\Phi_{33}
	\right) \rho \md \rho \md z .
\end{equation}
Note that there is a balancing of terms. The first term $\Phi_{00}$ is always positive, the second term is negative and grows with increasing $\omega$, while the third term is negative outside of an ergoregion, but becomes positive within an ergoregion. The net result is the behavior shown in \Figref{fig.SolutionSequences.HMVdeficit_vs_E0} and by the horizontal line in each plot of \Figref{fig.DeficitAngle}. It is interesting that extreme members along the string sequence have a deficit angle that grows to this value outside the support of the matter and before transitioning to the far-field regime. We note that similar behavior, including agreement of the deficit angle with the result of \Eqref{eq.LinearDiracDeficitEstimate}, occurs for the extreme members of the $L_0 = 0.95$ solution sequence.

\newcommand{\hy}{3cm}
\begin{figure*}[b!]
    \centering
    \begin{subfigure}[t]{0.3\textwidth}
        \centering
        \includegraphics[height=\hy]{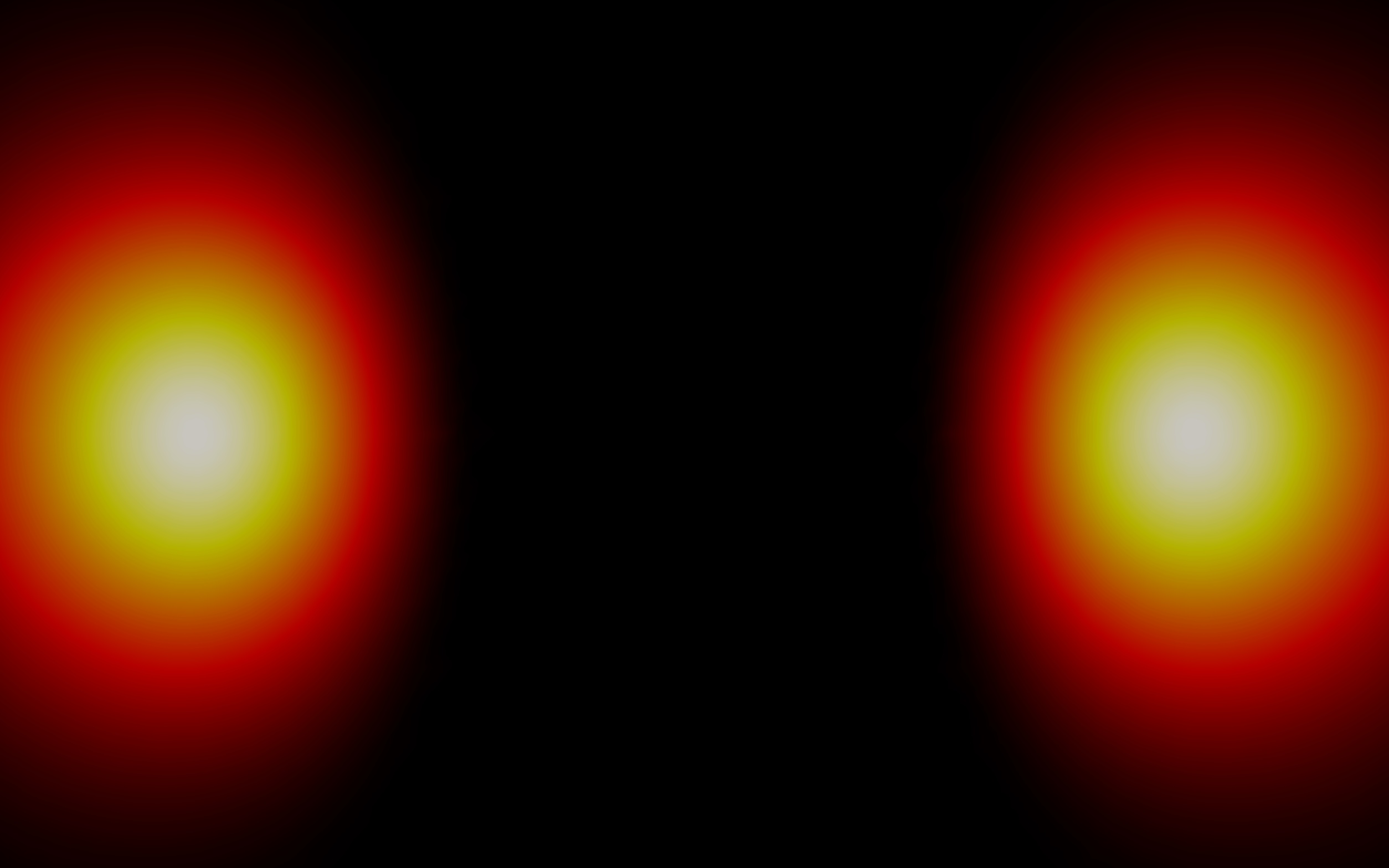}
        \caption{$E_0 = 0.8$}
    \end{subfigure}%
    ~
    \begin{subfigure}[t]{0.3\textwidth}
        \centering
        \includegraphics[height=\hy]{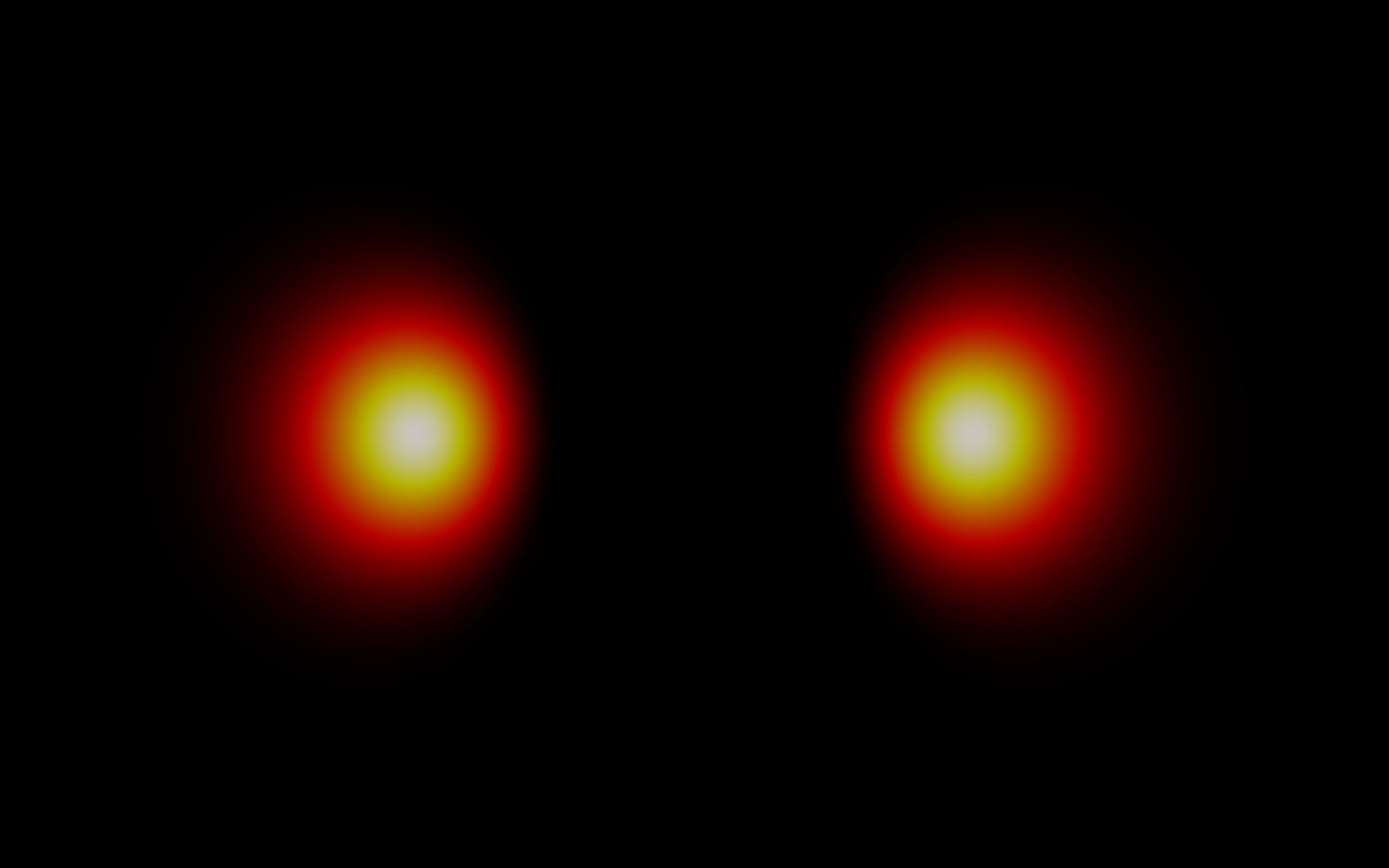}
        \caption{$E_0 = 0.7$}
    \end{subfigure}%
    ~
    \begin{subfigure}[t]{0.3\textwidth}
        \centering
        \includegraphics[height=\hy]{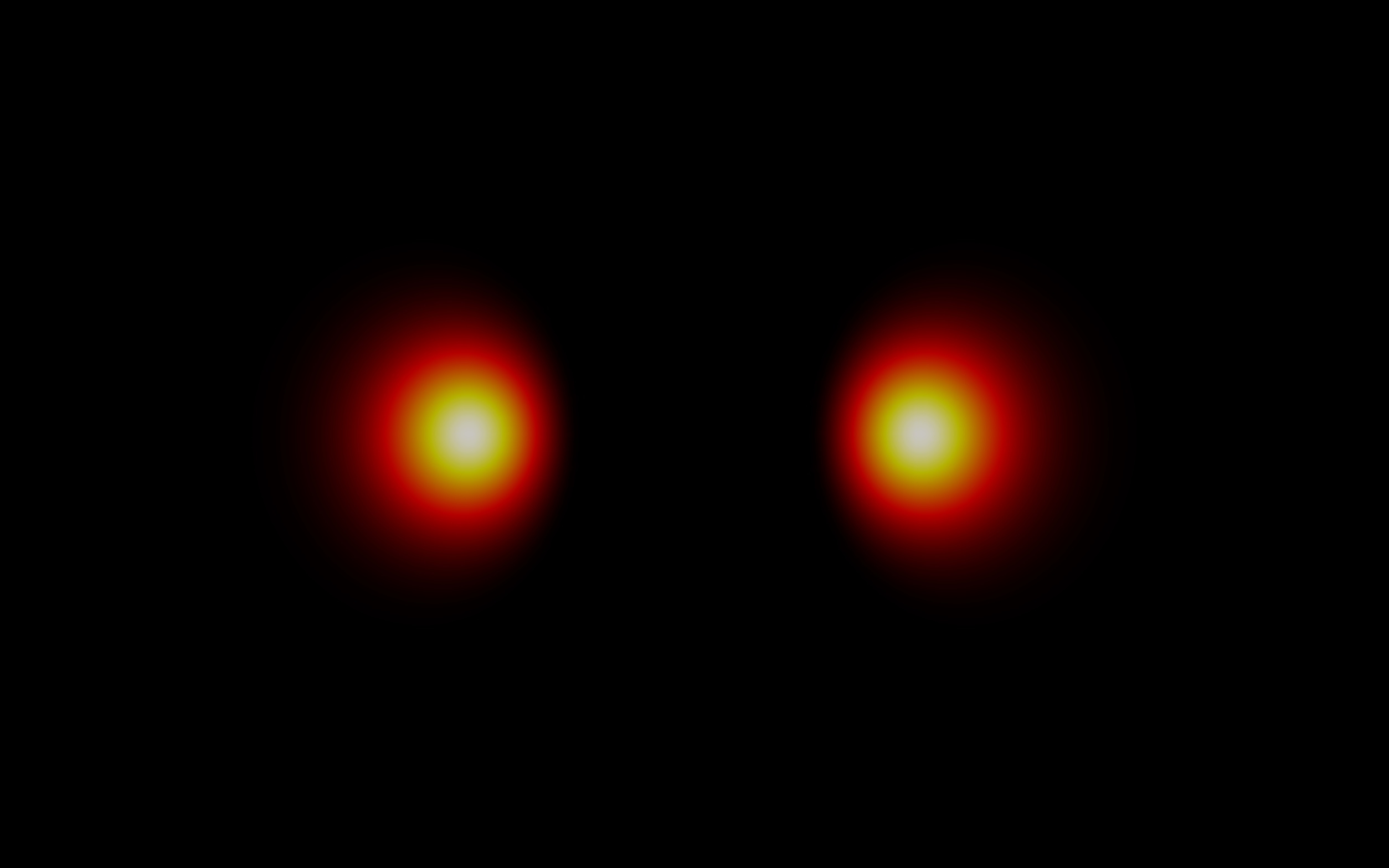}
        \caption{$E_0 = 0.66$}
    \end{subfigure}

    \vspace{0.2cm}

    \begin{subfigure}[t]{0.3\textwidth}
        \centering
        \includegraphics[height=\hy]{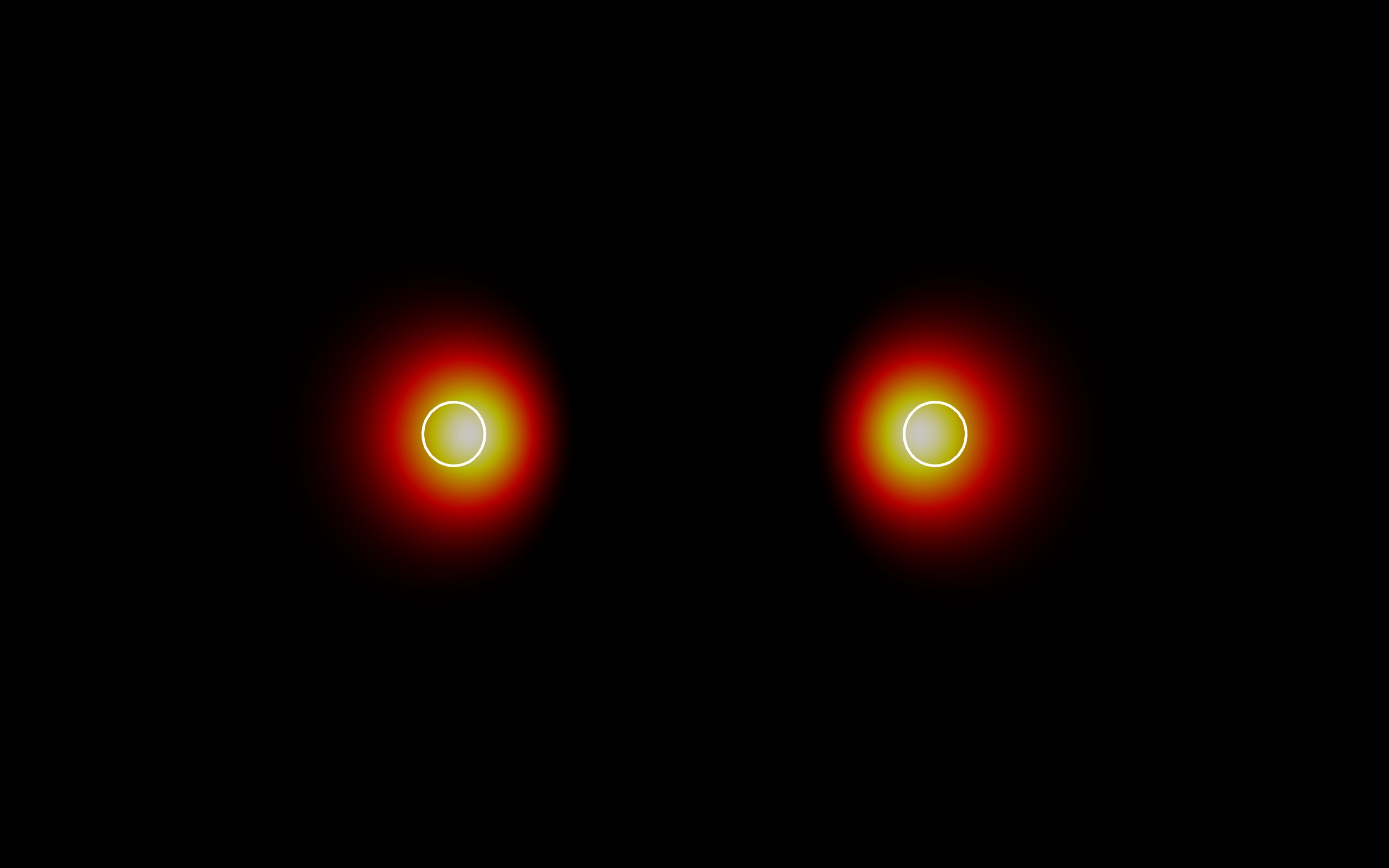}
        \caption{$E_0 = 0.65$}
    \end{subfigure}%
    ~
    \begin{subfigure}[t]{0.3\textwidth}
        \centering
        \includegraphics[height=\hy]{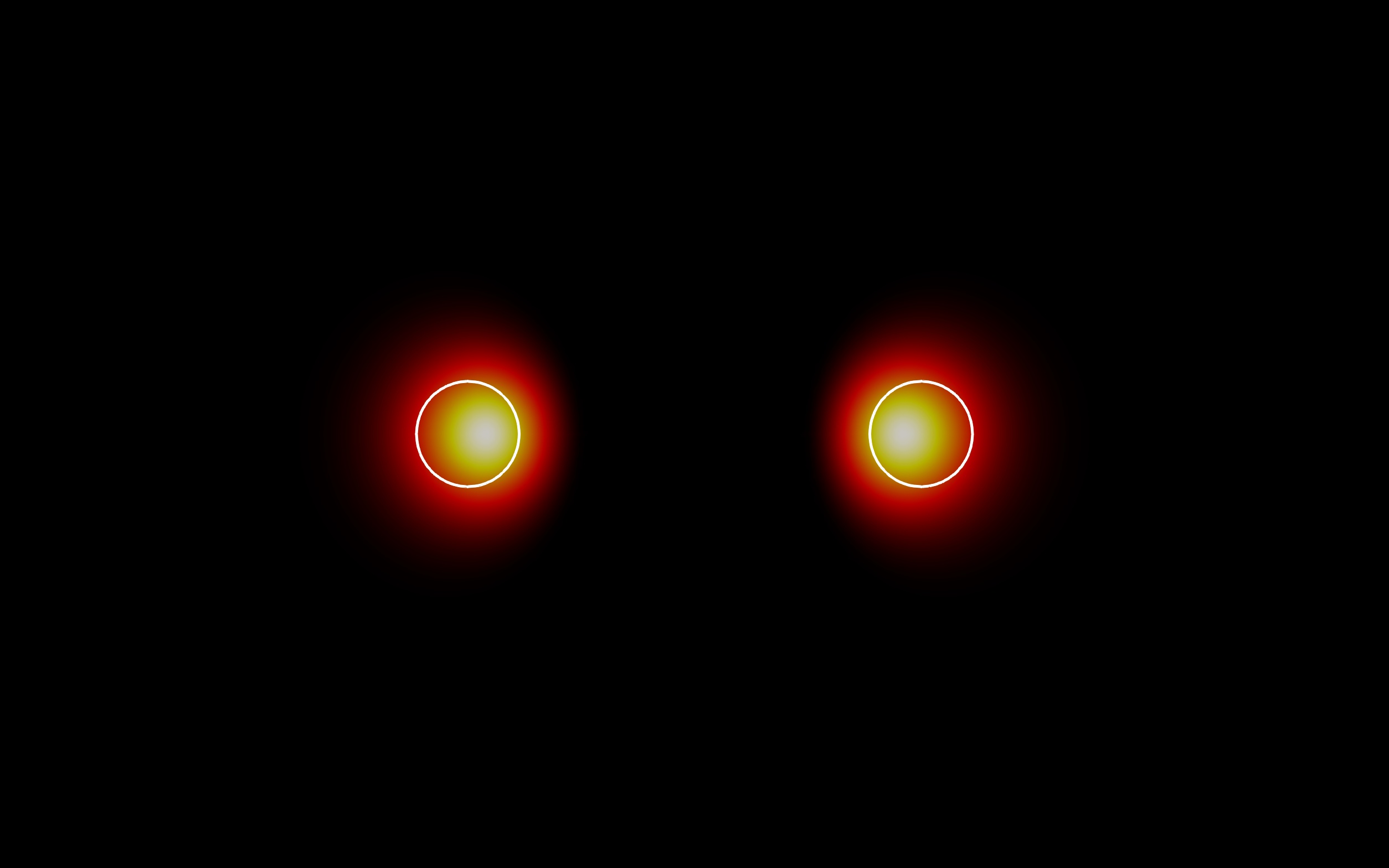}
        \caption{$E_0 = 0.64$}
    \end{subfigure}%
    ~
    \begin{subfigure}[t]{0.3\textwidth}
        \centering
        \includegraphics[height=\hy]{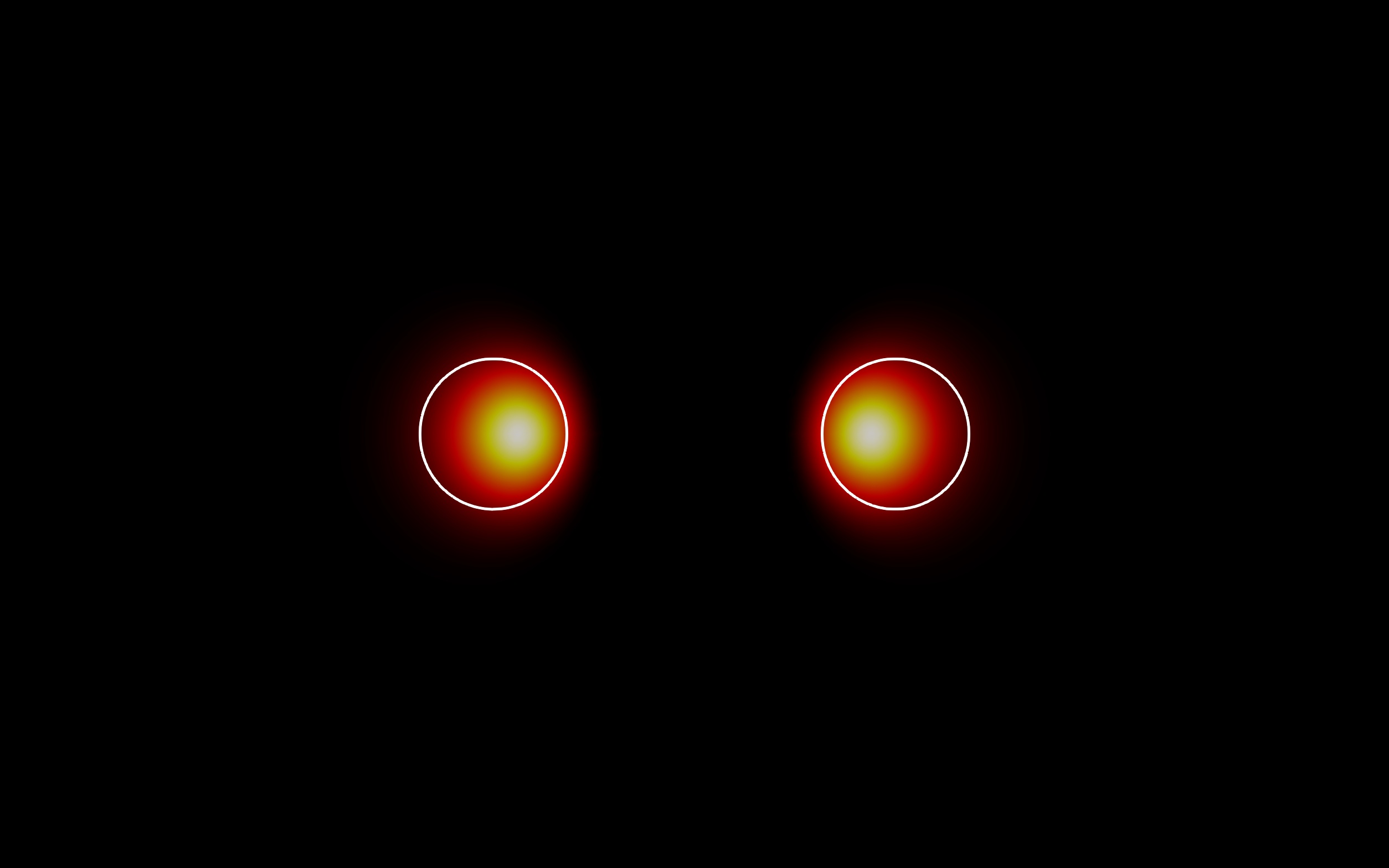}
        \caption{$E_0 = 0.625$}
    \end{subfigure}

    \vspace{0.2cm}

    \begin{subfigure}[t]{0.3\textwidth}
        \centering
        \includegraphics[height=\hy]{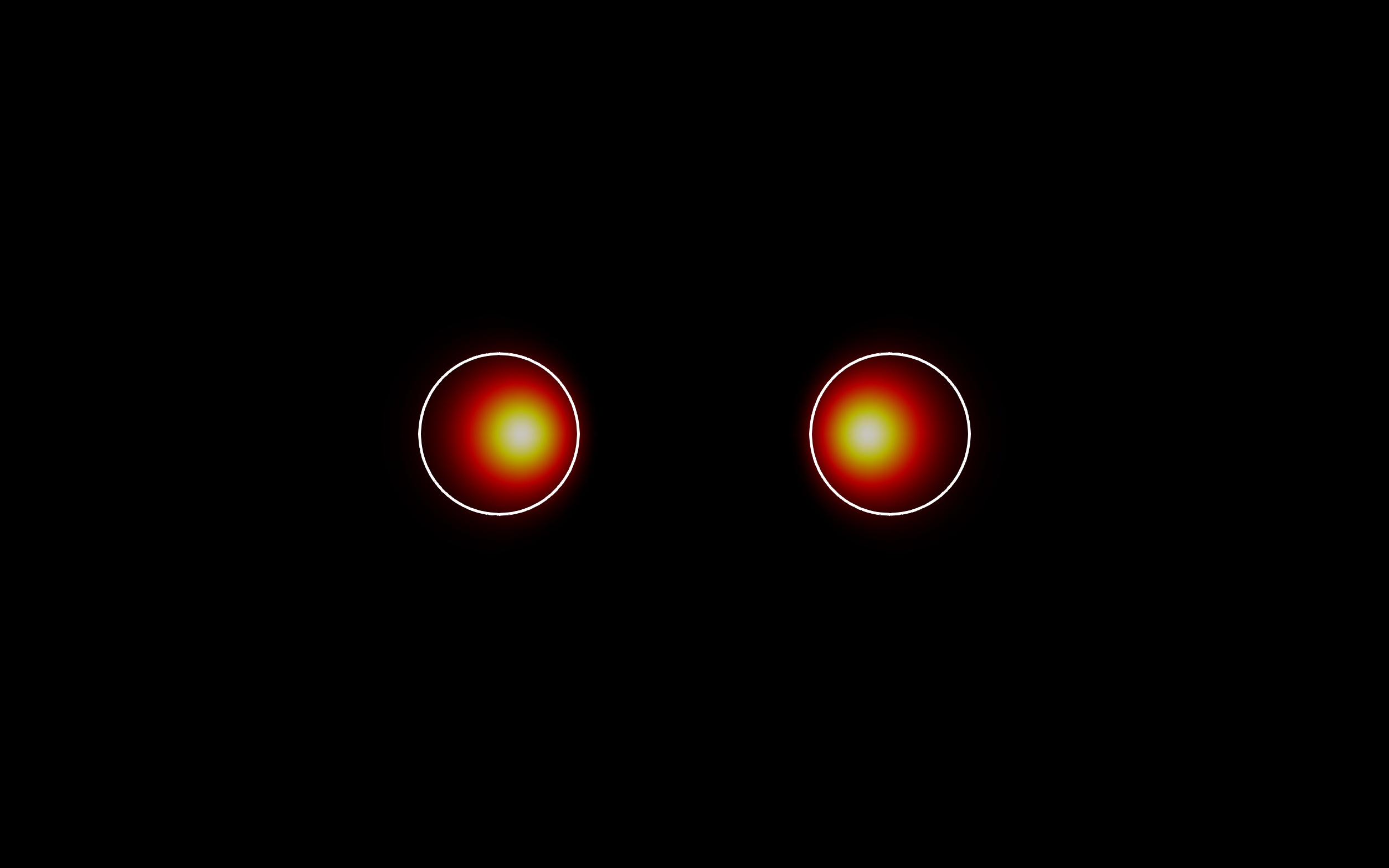}
        \caption{$E_0 = 0.6$}
    \end{subfigure}%
    ~
    \begin{subfigure}[t]{0.3\textwidth}
        \centering
        \includegraphics[height=\hy]{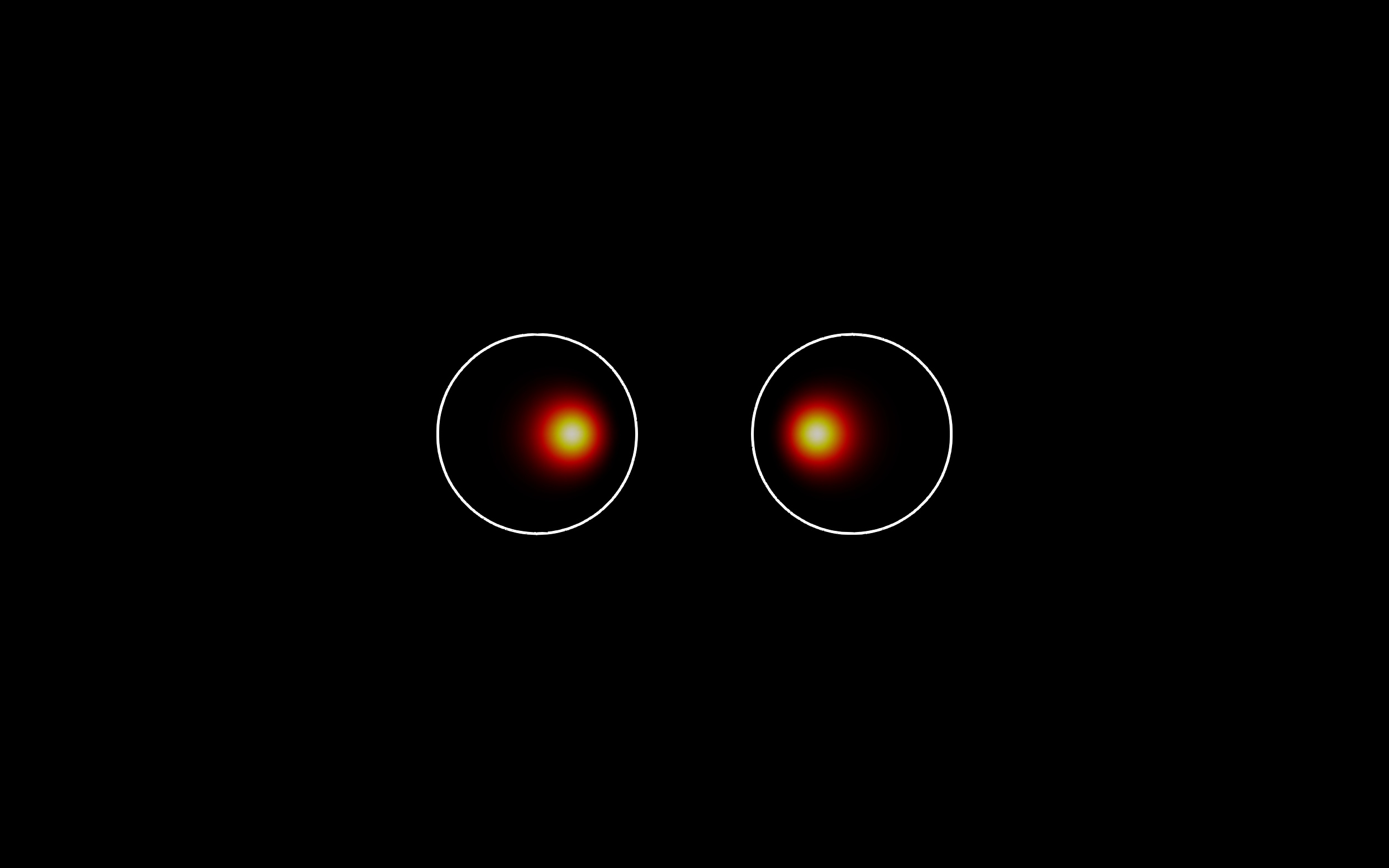}
        \caption{$E_0 = 0.54$}
    \end{subfigure}%
    ~
    \begin{subfigure}[t]{0.3\textwidth}
        \centering
        \includegraphics[height=\hy]{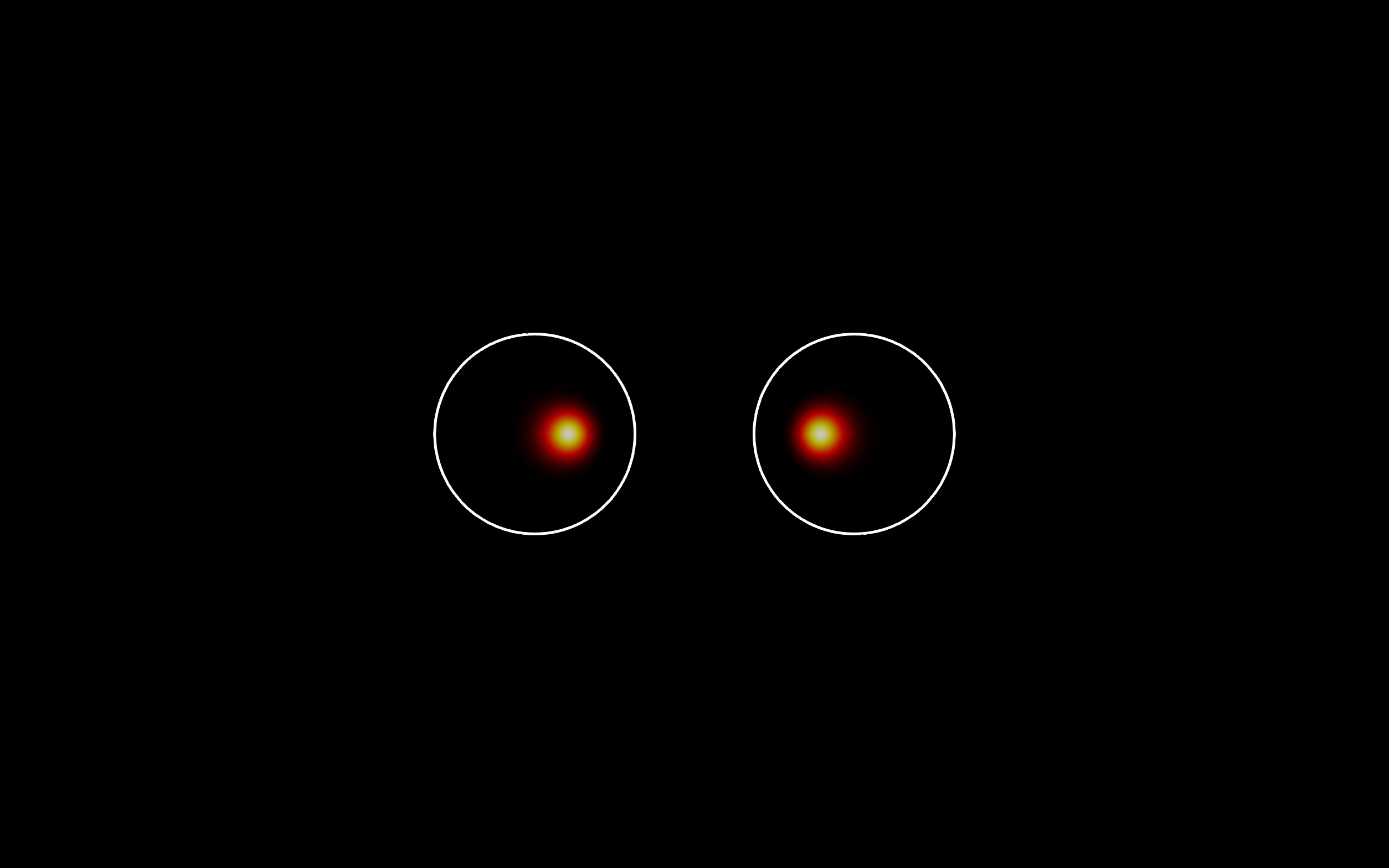}
        \caption{$E_0 = 0.52$}
    \end{subfigure}

    \vspace{0.2cm}

    \begin{subfigure}[t]{0.3\textwidth}
        \centering
        \includegraphics[height=\hy]{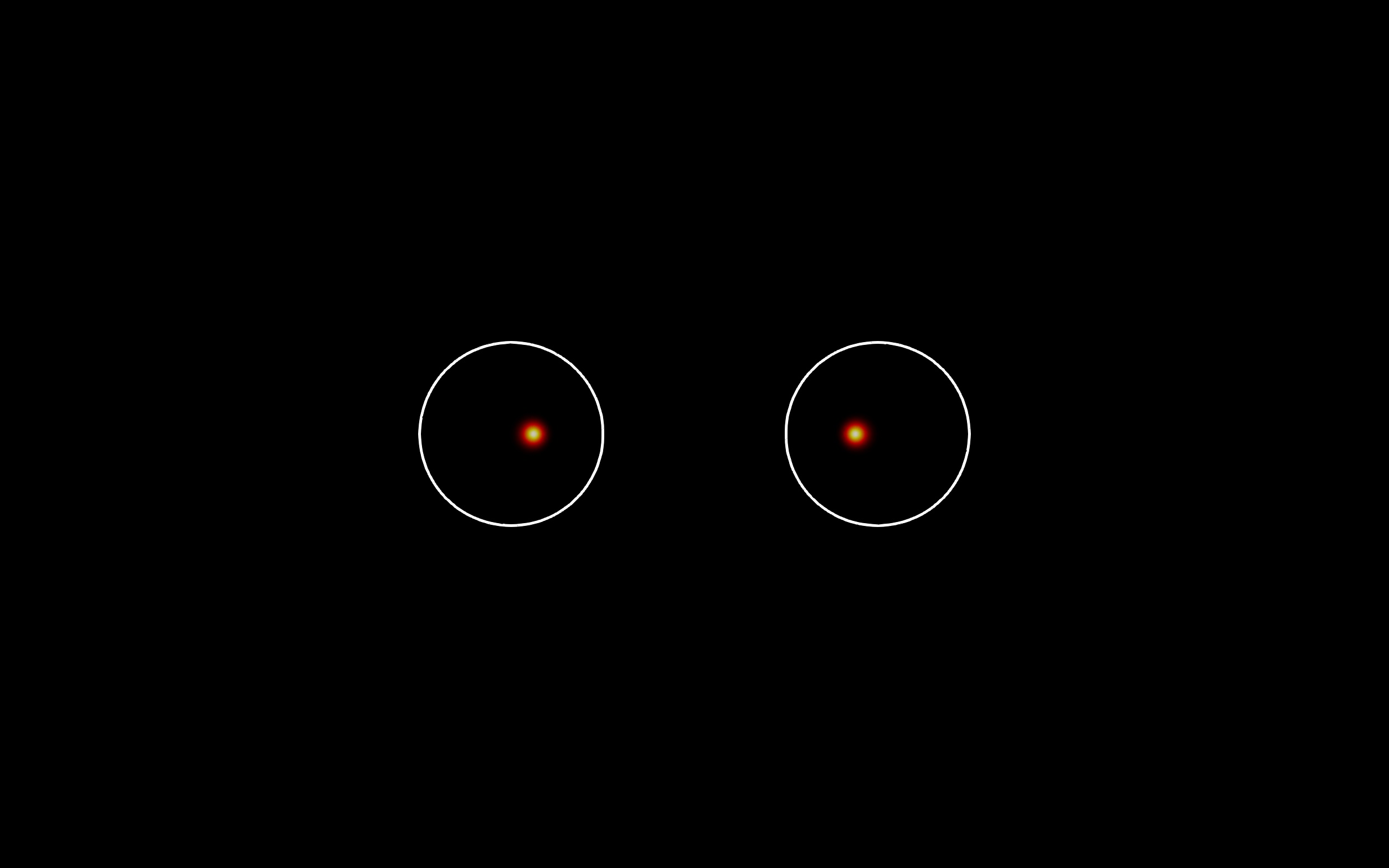}
        \caption{$E_0 = 0.5$}
    \end{subfigure}%
    ~
    \begin{subfigure}[t]{0.3\textwidth}
        \centering
        \includegraphics[height=\hy]{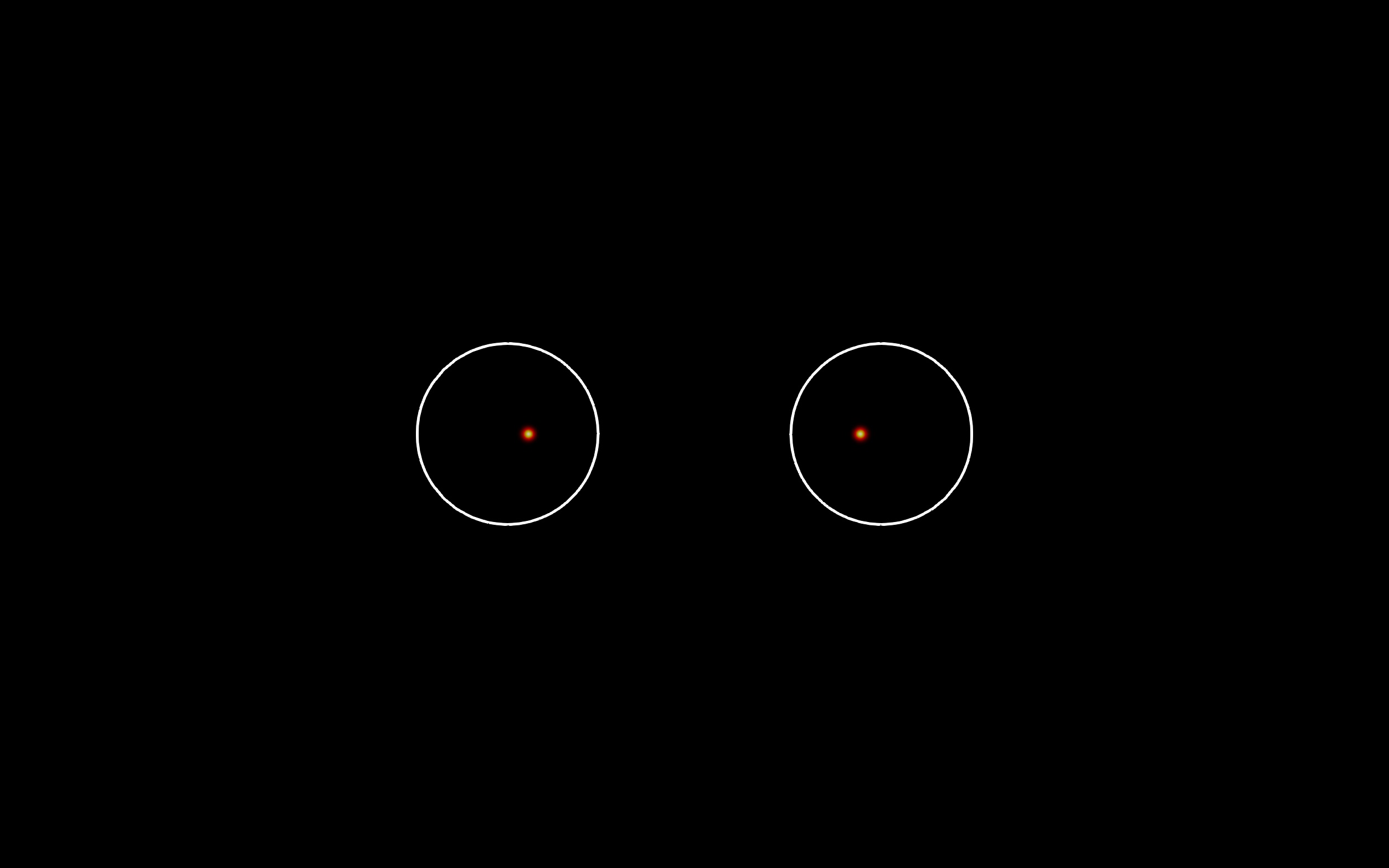}
        \caption{$E_0 = 0.485$}
    \end{subfigure}%
    ~
    \begin{subfigure}[t]{0.3\textwidth}
        \centering
        \includegraphics[height=\hy]{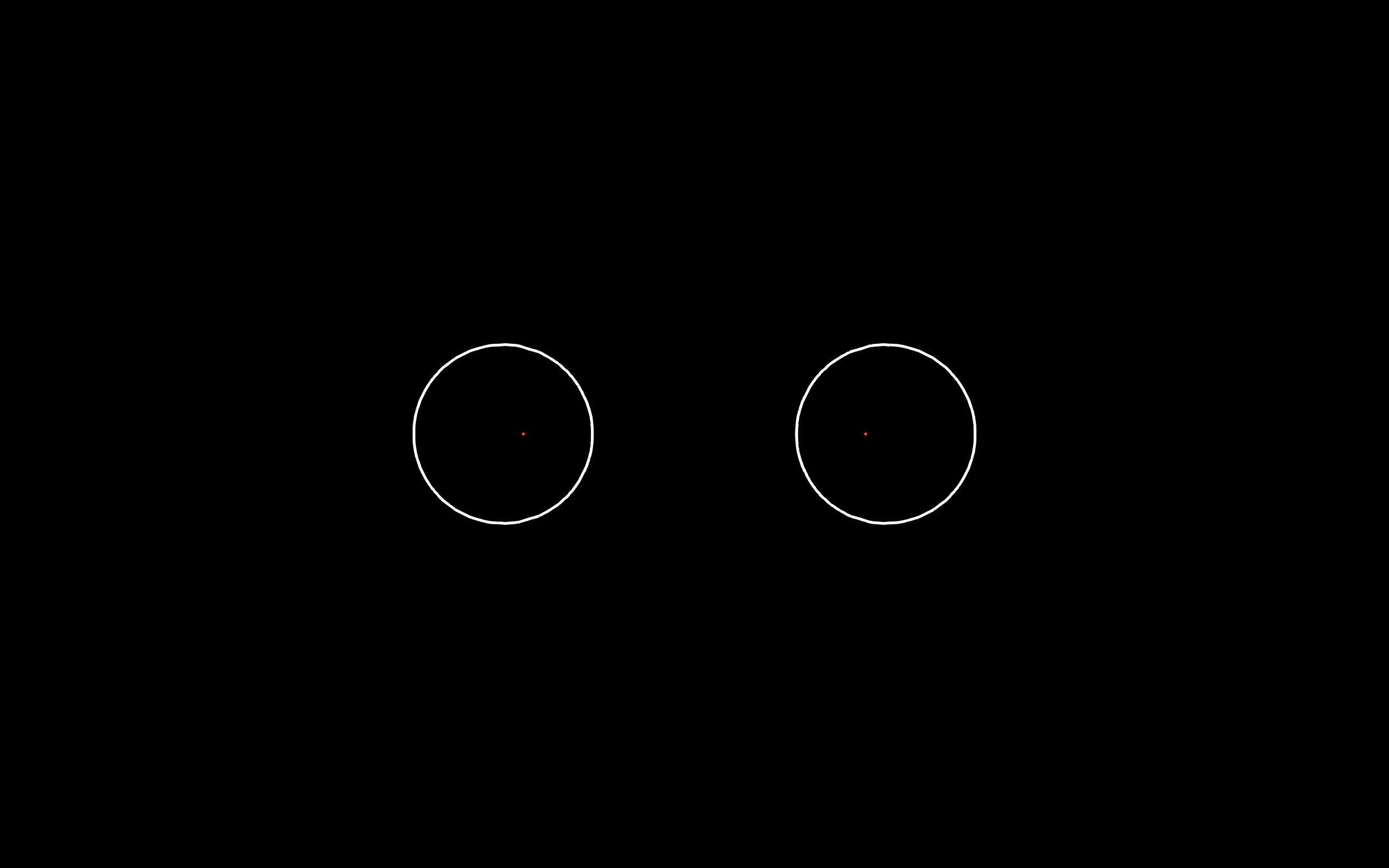}
        \caption{$E_0 = 0.46$}
    \end{subfigure}
    \caption{Energy density (heat map) and ergoregion (boundary shown by white trace) for a selection of solutions on the $L_0 = 0.8$ solution sequence.}
   \label{fig.RHOErgo2D}
\end{figure*}

\begin{figure}[tb!]
\label{fig.RingandErgo.3D}
    \centering
    \begin{subfigure}[t]{0.3\textwidth}
        \centering
        \includegraphics[height=\hy]{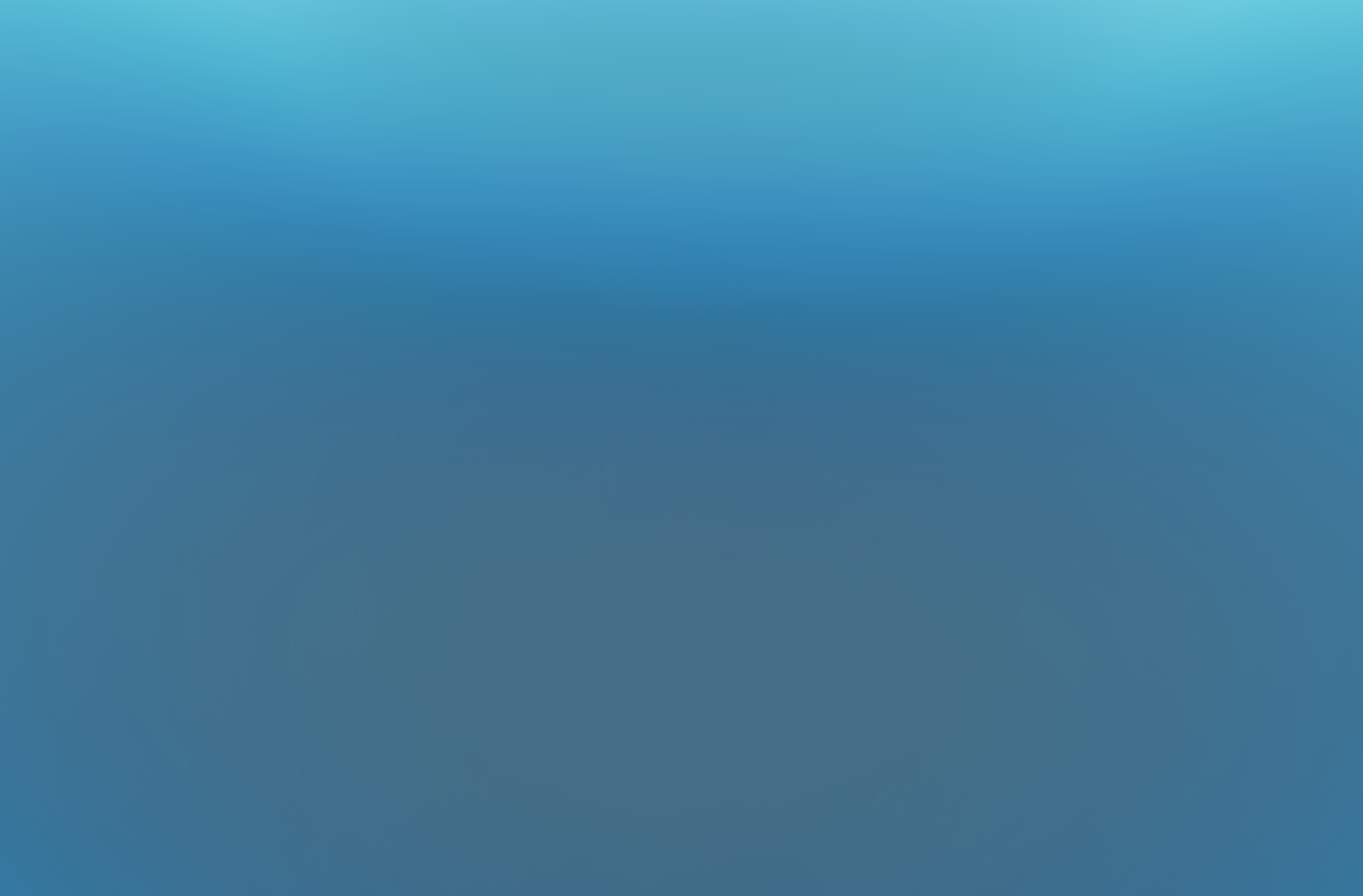}
        \caption{$E_0 = 0.8$}
    \end{subfigure}%
    ~
    \begin{subfigure}[t]{0.3\textwidth}
        \centering
        \includegraphics[height=\hy]{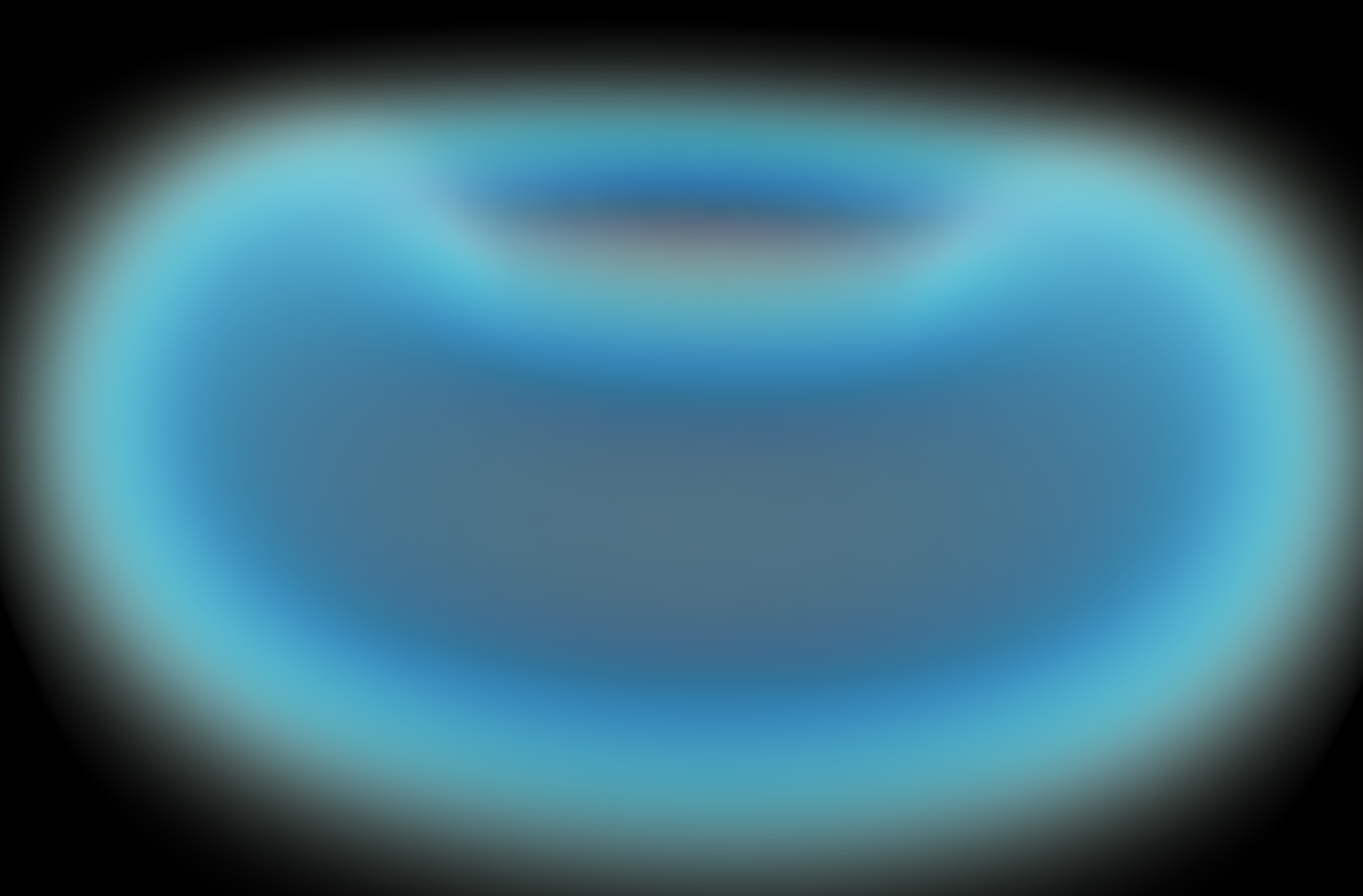}
        \caption{$E_0 = 0.7$}
    \end{subfigure}%
    ~
    \begin{subfigure}[t]{0.3\textwidth}
        \centering
        \includegraphics[height=\hy]{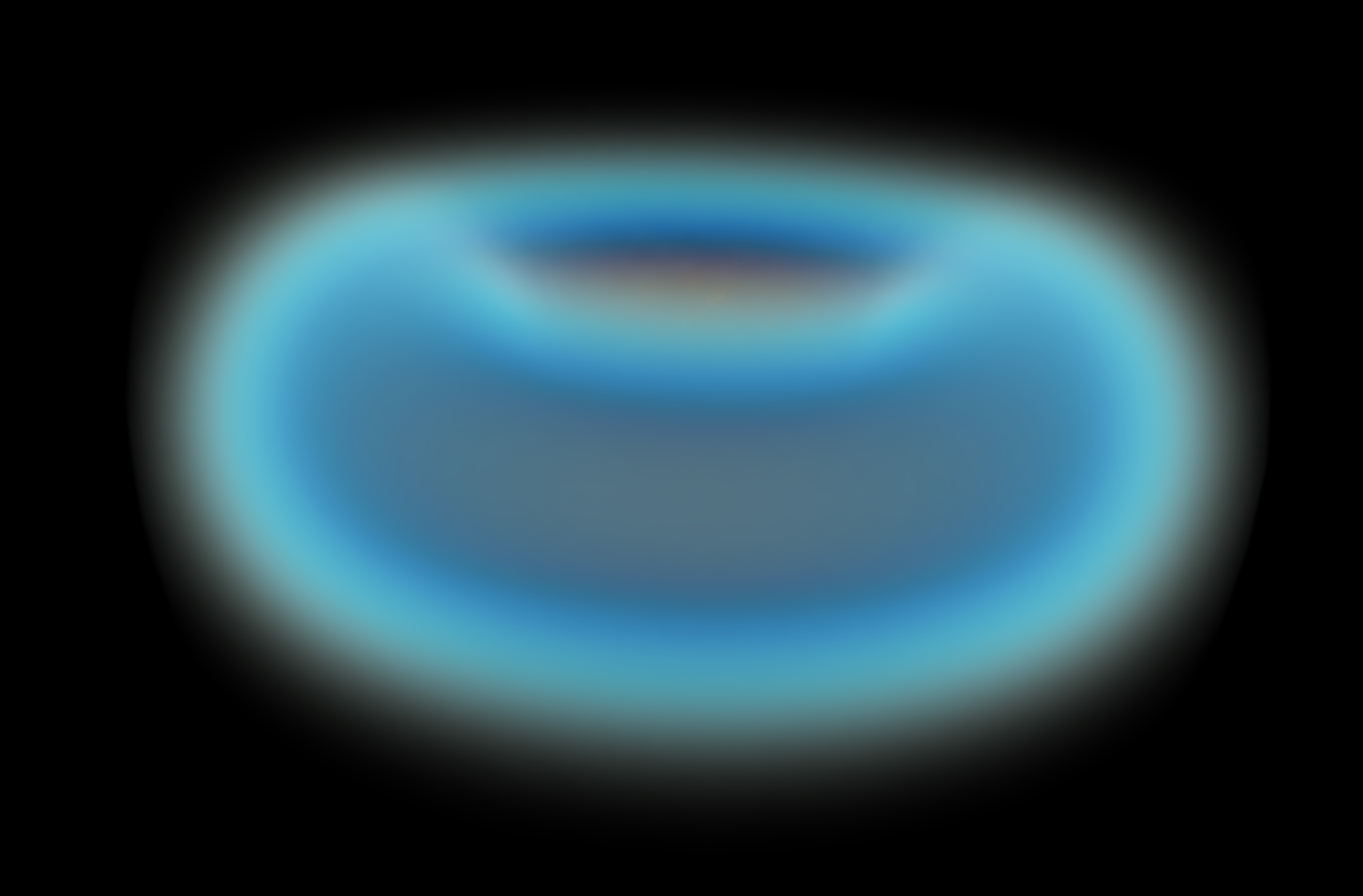}
        \caption{$E_0 = 0.66$}
    \end{subfigure}

    \vspace{0.2cm}

    \begin{subfigure}[t]{0.3\textwidth}
        \centering
        \includegraphics[height=\hy]{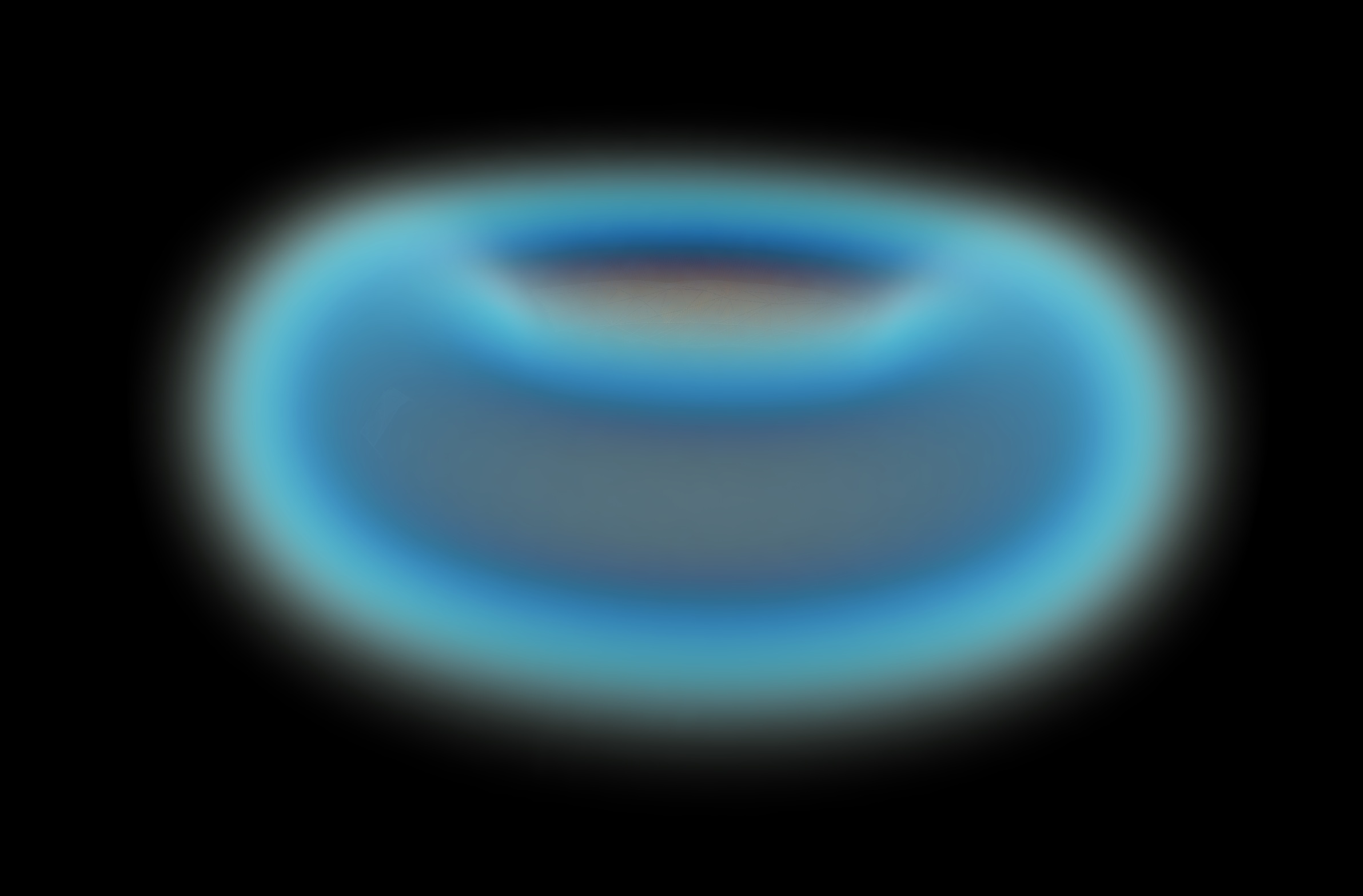}
        \caption{$E_0 = 0.65$}
    \end{subfigure}%
    ~
    \begin{subfigure}[t]{0.3\textwidth}
        \centering
        \includegraphics[height=\hy]{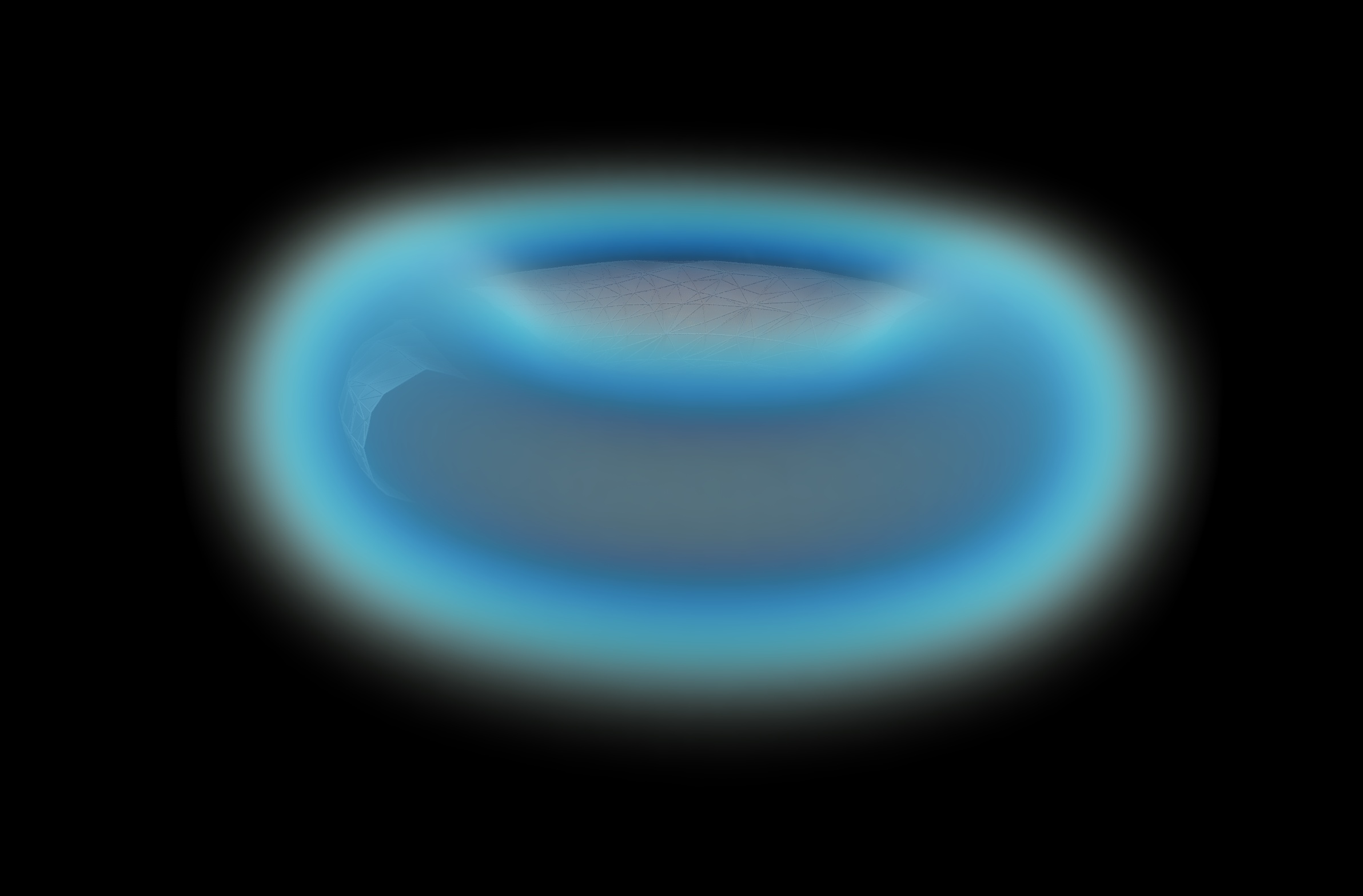}
        \caption{$E_0 = 0.64$}
    \end{subfigure}%
    ~
    \begin{subfigure}[t]{0.3\textwidth}
        \centering
        \includegraphics[height=\hy]{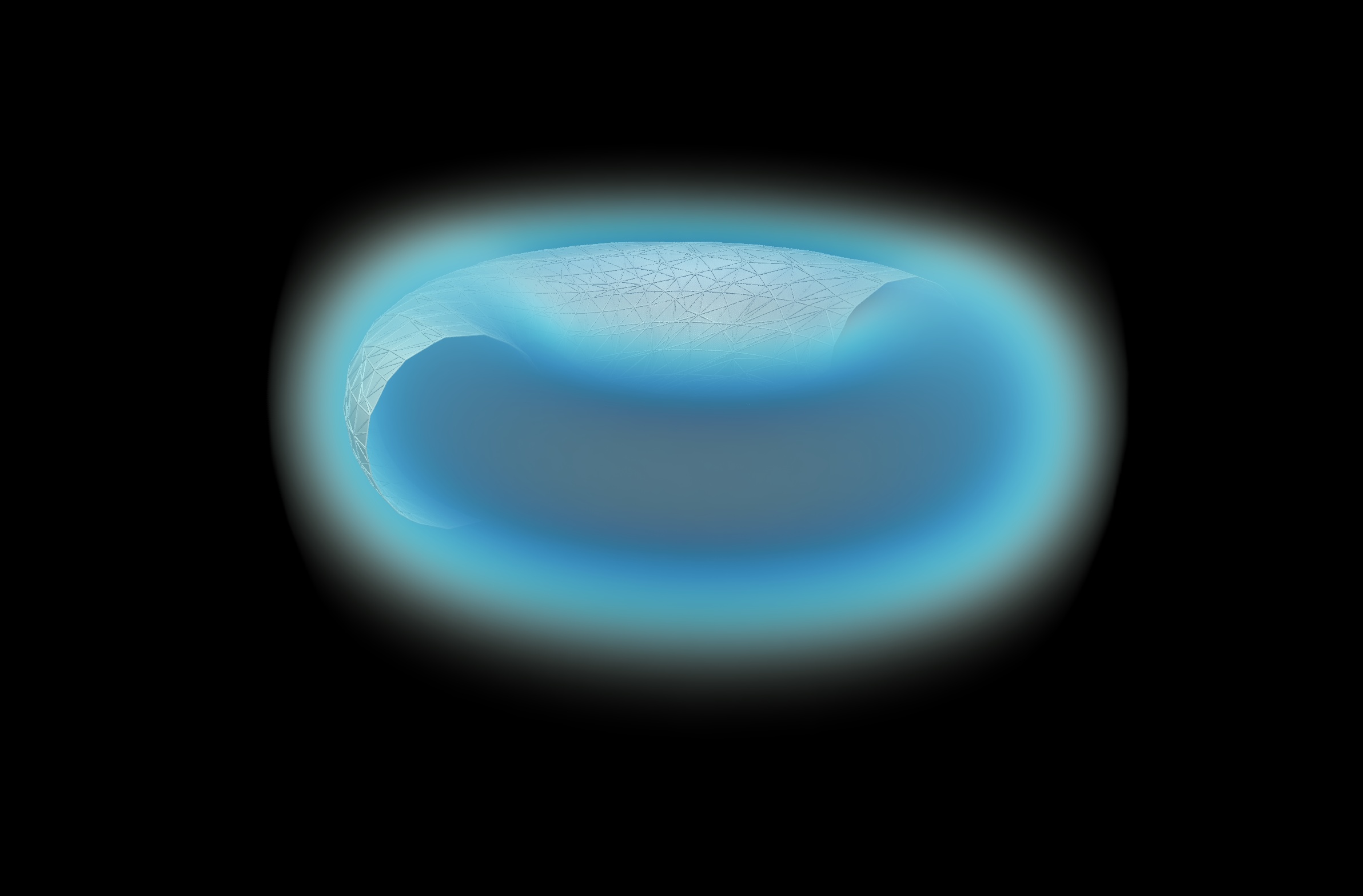}
        \caption{$E_0 = 0.625$}
    \end{subfigure}

    \vspace{0.2cm}

    \begin{subfigure}[t]{0.3\textwidth}
        \centering
        \includegraphics[height=\hy]{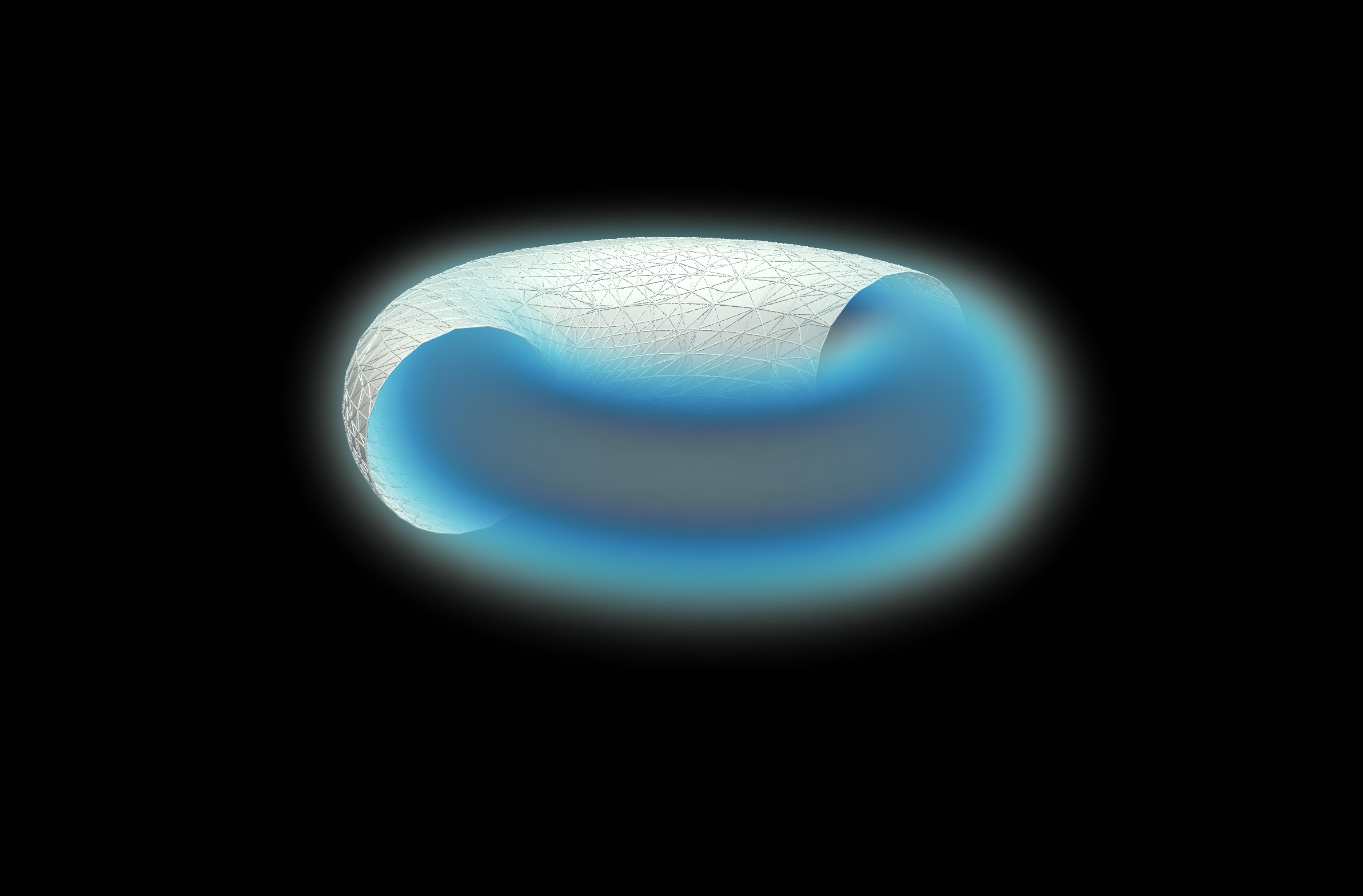}
        \caption{$E_0 = 0.6$}
    \end{subfigure}%
    ~
    \begin{subfigure}[t]{0.3\textwidth}
        \centering
        \includegraphics[height=\hy]{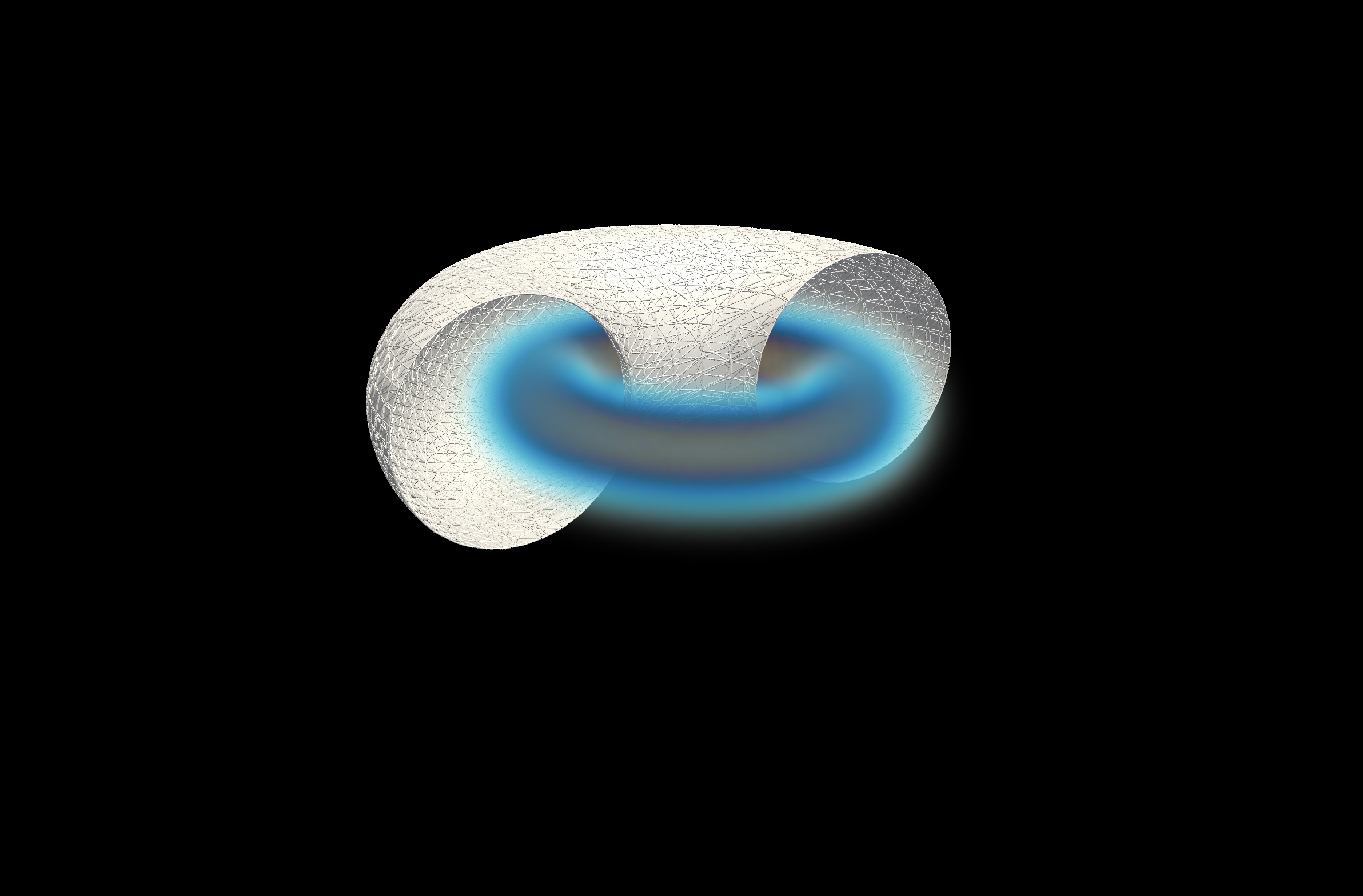}
        \caption{$E_0 = 0.54$}
    \end{subfigure}%
    ~
    \begin{subfigure}[t]{0.3\textwidth}
        \centering
        \includegraphics[height=\hy]{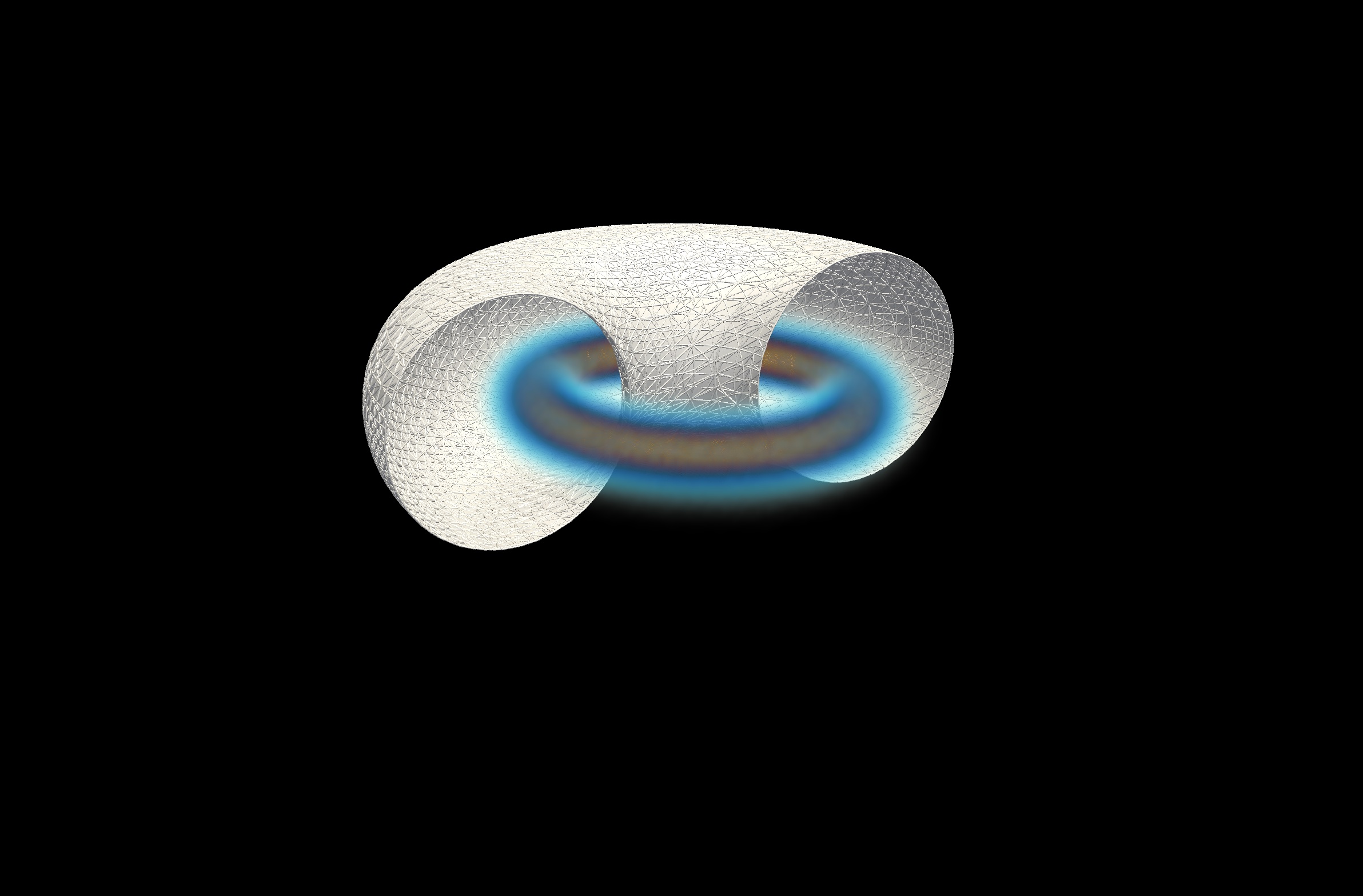}
        \caption{$E_0 = 0.52$}
    \end{subfigure}

    \vspace{0.2cm}

    \begin{subfigure}[t]{0.3\textwidth}
        \centering
        \includegraphics[height=\hy]{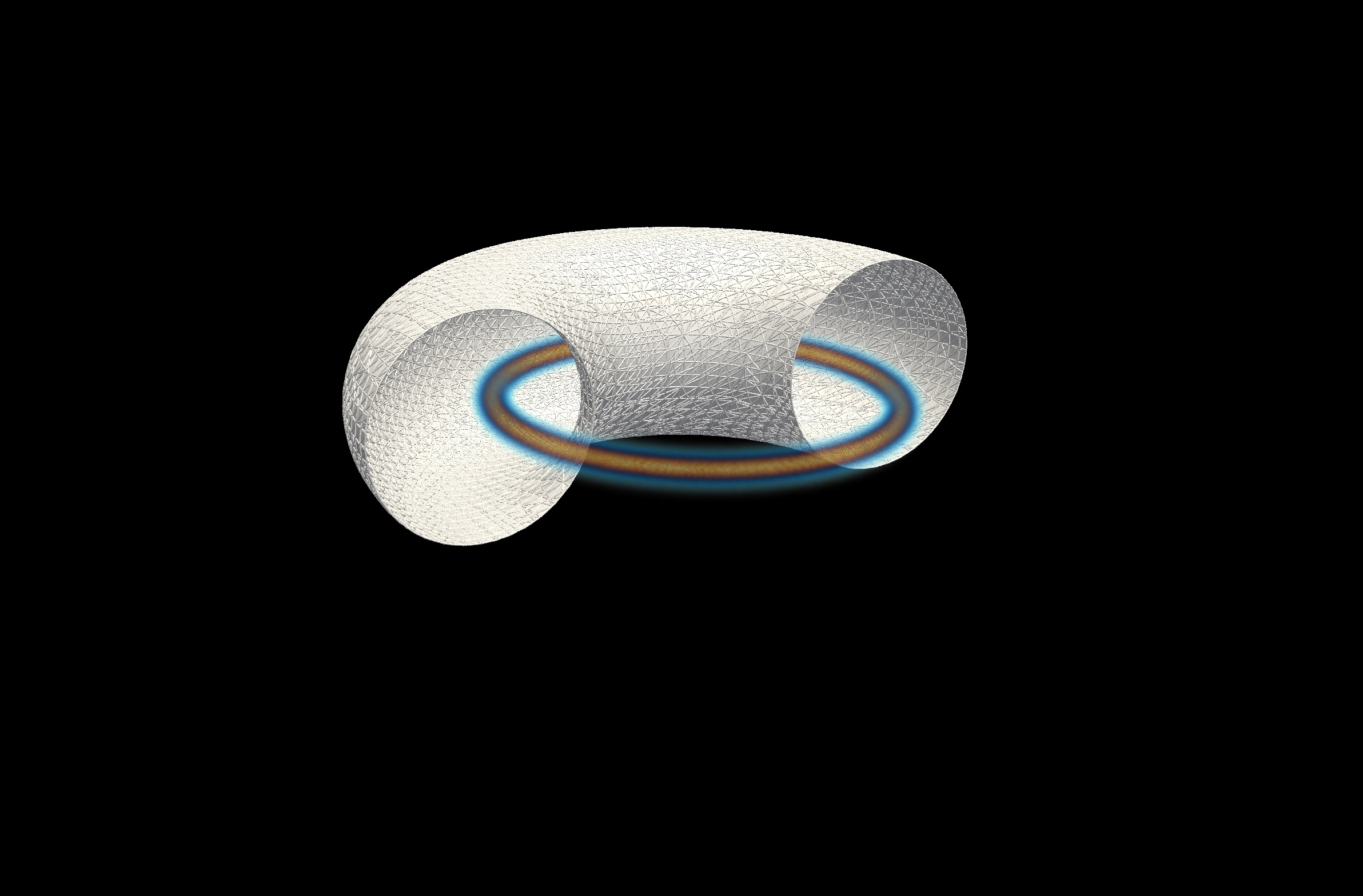}
        \caption{$E_0 = 0.5$}
    \end{subfigure}%
    ~
    \begin{subfigure}[t]{0.3\textwidth}
        \centering
        \includegraphics[height=\hy]{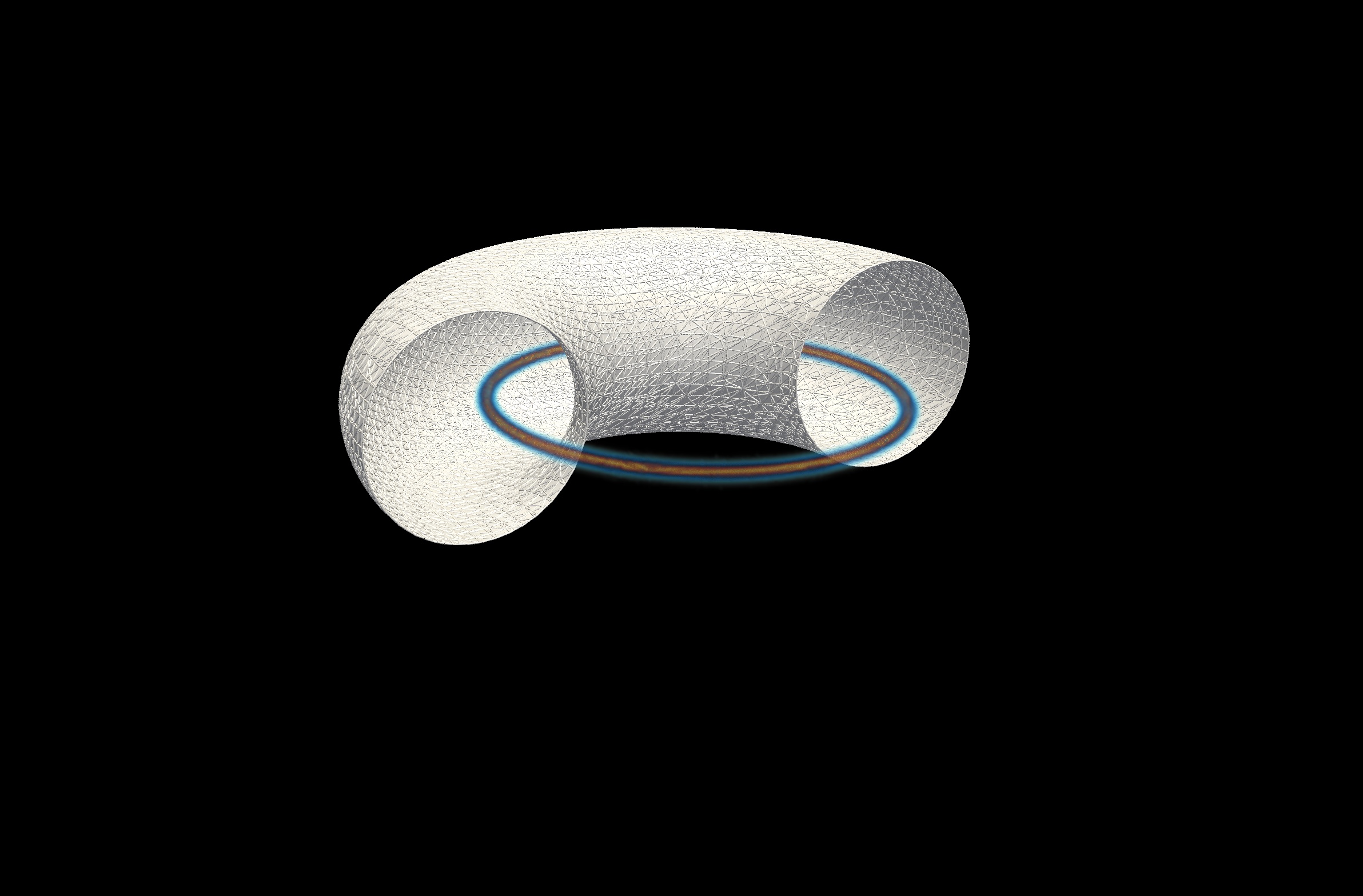}
        \caption{$E_0 = 0.485$}
    \end{subfigure}%
    ~
    \begin{subfigure}[t]{0.3\textwidth}
        \centering
        \includegraphics[height=\hy]{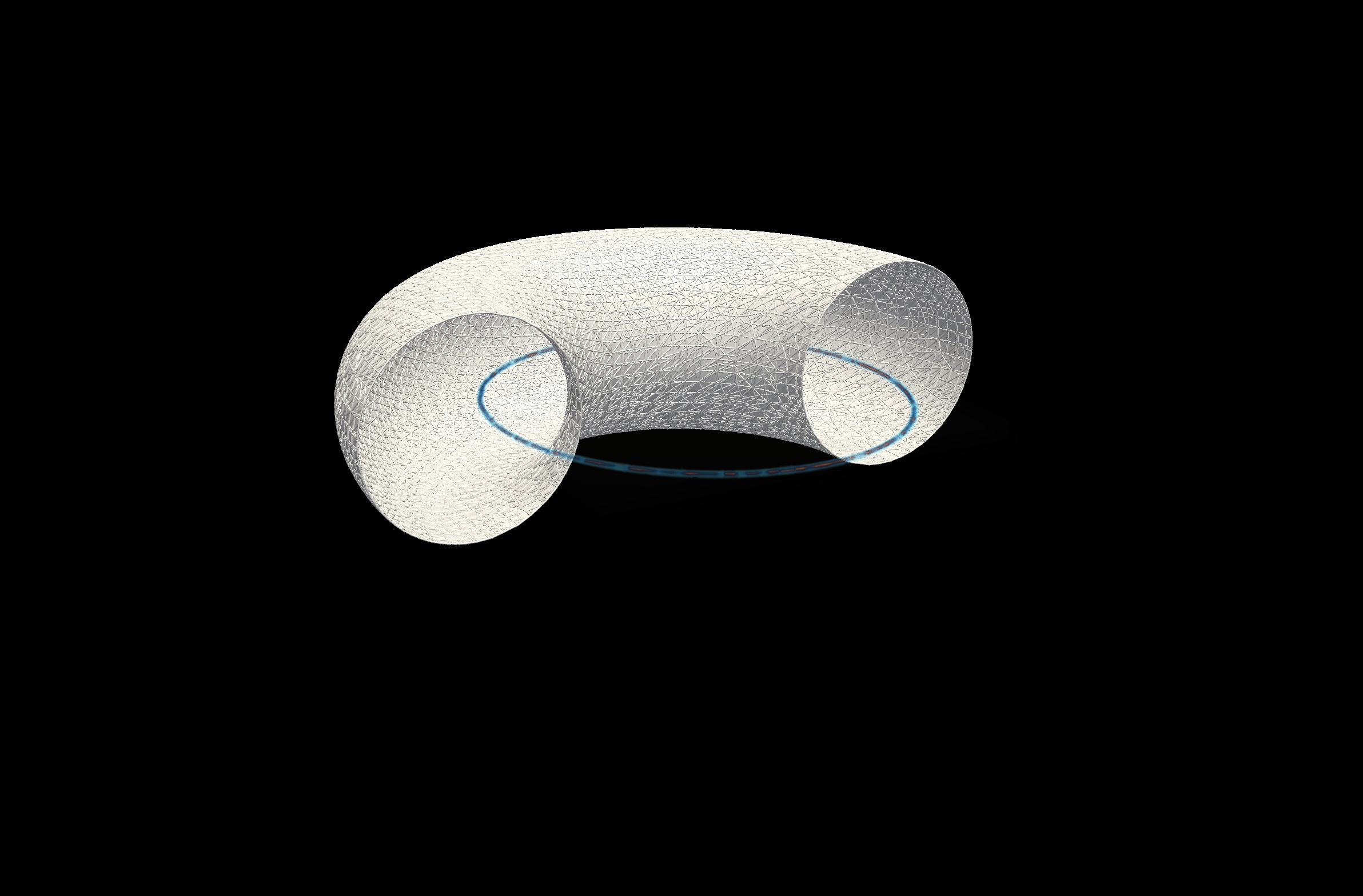}
        \caption{$E_0 = 0.46$}
    \end{subfigure}
    \caption{Energy density (blue) and ergoregion (surface shown in white) for a selection of solutions on the $L_0 = 0.8$ solution sequence.}
    \label{fig.RHOErgo3D}
\end{figure}

\section{Discussion and Conclusions}
\label{sec.DiscussionConlusions}
We have investigated a two-parameter family of toroidal stationary solutions to the Einstein-Vlasov system, and find evidence for two interesting physical limits. In \Sectionref{sec.BHlimit} we show that the Einstein-Vlasov system likely admits a quasistationary approach to an extremal Kerr black hole. This result is similar to that identified in axially symmetric spacetimes with uniformly rotating fluids \cite{Meinel:2006eh,Meinel:2004hj,Meinel:2012tn,Fischer:2005bw,Ansorg:2003dk}.
We also show in \Sectionref{sec.CosmicStringLimit} that high angular momentum solutions exhibit a distinct solution path, along which the peak in the ring density moves to larger radii and the geometry becomes locally conical about the ring. These solutions are compared with models for circular cosmic strings with Dirac sources, and appear consistent with the results of \cite{Frolov:1989ix,Hughes:1993ur,McManus:1993fp}. The limiting members of such sequences could therefore provide fully self-consistent models for circular cosmic strings. While Garfinkle and coauthors have considered fully self-consistent models in the case of straight strings, we are unaware of similar studies in the circular case. Kunz et al. \cite{Kunz:2013ju} have numerically modeled gravitating toroidal configurations of $U(1)\times U(1)$ gauge-fields (so-called \emph{vortons}), but they do not discuss a deficit angle.

To what extent can the limits of these solution sequences be pushed further? In the cosmic string limit our computation is eventually resolution limited even with mesh refinement. The matter becomes unresolved, and lost by the code resulting in a ``zero mass distribution'' error and code breakdown. It is likely that the limiting solution along such sequences has a Dirac distribution with radius and terminating $E_0$ value depending on the $L_0$ value. We believe that our solution sequences could be pushed further towards this limit by increasing resolution. While it is unlikely that such relativistic solutions are stable in full general relativity (without symmetry), their large angular momentum makes them stable against black hole collapse when restricted to axisymmetry.
With respect to the black hole limit we believe that we could go closer to the limiting solutions with a different numerical scheme. This is motivated by results in spherical symmetry where other methods allow one to numerically construct extremely thin shell solutions which are likely unstable \cite{Andreasson:2006dza}.

Finally we briefly comment on the potential physical relevance of these solutions. We have solved the equations with particle mass $m$ total mass $\mathcal M$ both equal to one. To get back a solution in physical units one may choose a value for $\mathcal M$ and rescale for instance the radius $r \to \mathcal M r$ and total angular momentum by $\mathcal J \to \mathcal M^2 \mathcal J$ (see for example \cite{Shapiro:1993hi}). There are no bounds on the length scale associated to the solutions presented in this paper, and thus under such a rescaling the solutions may represent objects from very small to astrophysical scales.

All of the near-extremal solutions (either in the black hole limit or in the string limit) contain ergoregions. Compact but non-black hole objects containing ergoregions are considered to be unstable and short-lived \cite{Friedman:1978tk,Cardoso:2008fh}. While stability in the Einstein-Vlasov system is largely wide-open, especially beyond spherical symmetry, the preference of our numerical scheme to find stable solutions and the instability results for high angular momentum solutions cited above indicate that solutions near both limits addressed in this paper are likely unstable.
Numerical experiments to investigate this issue are underway \cite{Ames:h5rQEYmq}.

\section*{Acknowledgements}
The authors thank Reinhard Meinel for useful comments during completion of this work. 
EA is supported by the Knut and Alice Wallenberg Foundation.

\bibliography{bibliography}
\end{document}